\documentclass[reprint,twocolumn,aps,prb,superscriptaddress]{revtex4-2}
\usepackage[T1]{fontenc}
\usepackage[utf8]{inputenc}
\usepackage{lmodern}
\usepackage{graphicx}
\usepackage{amsmath}
\usepackage{xcolor}
\usepackage{float}
\usepackage{hyperref}
\usepackage{braket}
\usepackage{amssymb,amsmath,amsthm, array}
\usepackage{mdframed}
\usepackage{systeme,mathtools}
\usepackage{multirow}

\DeclareUnicodeCharacter{0301}{-}

\begin{document}
\renewcommand{\vec}[1]{\mathbf{#1}}
\newcommand{\ii}{\mathrm{i}}
\def\ya#1{{\color{orange}{#1}}}

\title{Magnetoelectric fractals, Magnetoelectric parametric resonance and Hopf bifurcation}

\author{M. Wanic}
\affiliation{Department of Physics and Medical Engineering, Rzesz\'ow University of Technology, 35-959 Rzesz\'ow, Poland}

\author{Z. Toklikishvili}
\affiliation{Faculty of Exact and Natural Sciences, Tbilisi State University, Chavchavadze av.3, 0128 Tbilisi, Georgia}

\author{S. K. Mishra}

\affiliation{Department of Physics, Indian Institute of Technology (Banaras Hindu University) Varanasi - 221005, India}

\author{M. Trybus}
\affiliation{Department of Physics and Medical Engineering, Rzesz\'ow University of Technology, 35-959 Rzesz\'ow, Poland}

\author{L. Chotorlishvili}
\affiliation{Department of Physics and Medical Engineering, Rzesz\'ow University of Technology, 35-959 Rzesz\'ow, Poland}

\date{\today}
\begin{abstract}
In the present work, we study the dynamics of a magnetic nanoparticle coupled through the magnetoelectric coupling to the ferroelectric crystal. The model of our interest is nonlinear, and we explore the problem under different limits of weak and strong linearity. By applying two electric fields with different frequencies, we control the form of the confinement potential of the ferroelectric subsystem and realize different types of dynamics. We proved that the system is more sensitive to magnetoelectric coupling in the case of double-well potential. In particular, in the case of strong nonlinearity, arbitrary small values of magnetoelectric coupling lead to chaotic dynamics. In essence, magnetoelectric coupling plays a role akin to the small perturbations destroying invariant tors according to the KAM theorem. We showed that bifurcations in the system are of Hopf's type. We observed the formation of magnetoelectric fractals in the system. In the limit of weak nonlinearity, we studied a problem of parametric nonlinear resonance and enhancement of magnetic oscillations through magnetoelectric coupling. 
\end{abstract}

\maketitle

\section{Introduction}

The phenomenological approach to magnetism deals with the celebrated Landau–Lifshitz–Gilbert (LLG) equation. The LLG equation is a nonlinear equation that describes the dynamics of the magnetic moment in the effective magnetic field. For different effective magnetic fields, the dynamics of the magnetic moment can be diverse, from linear oscillations to soliton solutions or chaotic behavior in particular cases. The standard recipe of applying the LLG equation implies a coarse-graining procedure of the sample into the unit cells of the size of several nanometers (this procedure concerns both bulk or thin magnetic films). Consequently, for modeling  experimentally relevant physical phenomena, a set of coupled equations should be solved numerically. Together with the finite temperature effects, this restricts the application of analytic methods. Nevertheless, the advent and development of biologically inspired nanomaterials \cite{farzin2020magnetic,gloag2019advances,shasha2021nonequilibrium,liu2020preparation,wu2019magnetic}  entirely changed the agenda. The single magnetic moments of individual magnetic nanoparticles can be used in magnetic resonance imaging of biological textures, sensing, and diverse applications in biomedicine \cite{PhysRevLett.126.224501}. Studying the nonlinear dynamics of a single magnetic moment became an experimentally relevant task.

Heterostructures of two different subsystems possessing two different order parameters are highly interesting. The interface effect and coupling between two materials are essential in composite bilayer systems. A coupling between the ferroelectric and ferromagnetic materials is named magnetoelectric (ME) effect \cite{PhysRevB.107.115126,PhysRevB.91.041408,khomeriki2016positive}. In particular, we are interested in the ME coupling between the magnetic nanoparticle and ferroelectric single crystal. In this context, we admit an interesting phenomenon discovered recently, the magnetochiral anisotropy of Triglycine sulfate (TGS)  \cite{choudhury2004ferroelectric,choudhury2009structural,choudhury2003role,trybus2020phase,trybus2016observation,trybus2015dynamic}. The experimentally observed voltage ratio for a TGS crystal shows dependence on the magnetic field \cite{PhysRevB.106.224307,rikken2022dielectric} meaning the coupling between ferroelectric and magnetic properties in TGS. According to Ginzburg–Landau theory, ferroelectric systems are described by the phenomenological potential of the form $F_f(x,T)=\frac{p_x^2}{2m}+a(T-T_c)x^2+bx^4-xE$. Below the phase transition temperature $T<T_c$, phenomenological potential $F_f(x,T)$ has two minimum points, and when an external electric field is zero $E=0$, the ground state is double degenerated. 
We note that the biharmonic phenomenological potential is widely used for ferroelectric materials \cite{PhysRevLett.111.117202}. A prototypical ferroelectric system could also be BaTiO$_3$ in the
tetragonal phase \cite{PhysRevB.92.134424}. For BaTiO$_3$ system, the ME coupling is well studied experimentally \cite{PhysRevB.81.144425,PhysRevB.76.092108,PhysRevLett.97.047201}. ME coupling term couples the magnetic $m_z$ component of the magnetization vector with the ferroelectric order parameter as follows $V_{ME}=-g_{ME}xm_z$. Through the ME coupling we couple single ferroelectric crystal with the magnetic nanoparticle \cite{sharma2017synthesis} and describe the magnetic subsystem through a single coarse-grained magnetic moment. 

\begin{figure*}[!ht]
\centering
\includegraphics[width=\textwidth]{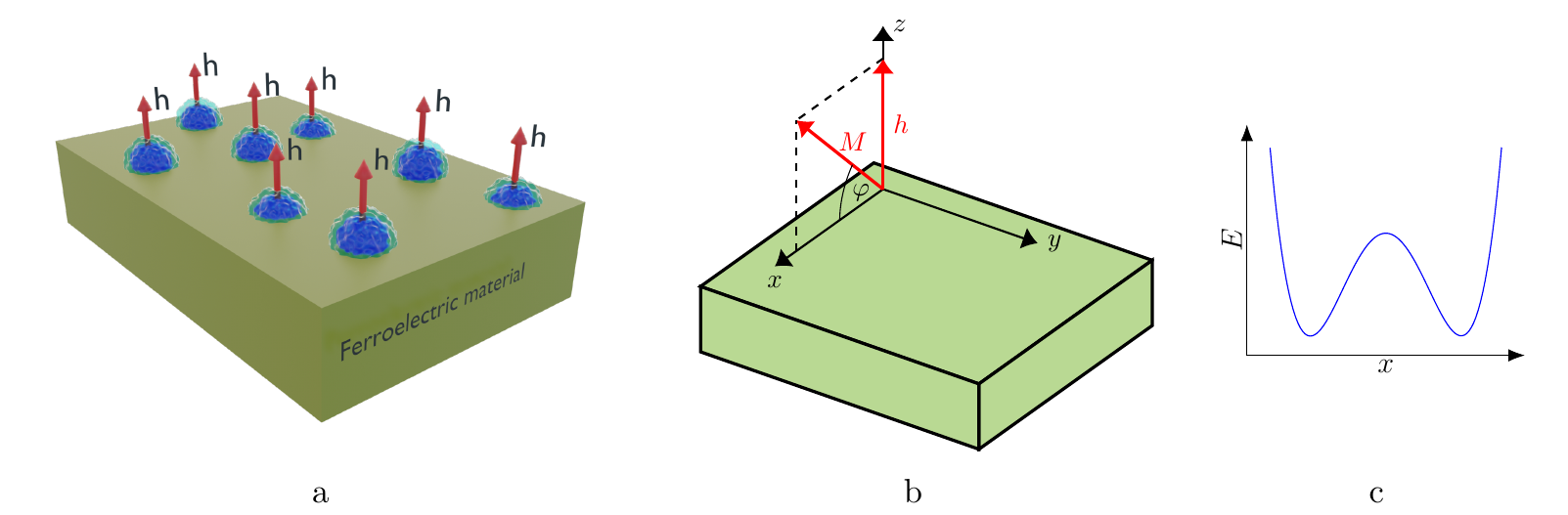}
\caption{Schematics of the composite structure under study:
Single-domain magnetic nanoparticles are deposited on a ferroelectric substrate (a). The screening effect leads to magneto-electric coupling between the ferroelectric dipole moment and magnetic nanoparticle. The interface polarization $P$ is driven by applied eternal electric fields. A constant external magnetic field is applied along the $Z$ axis. Dynamics of the magnetic moment $M$ of the magnetic nanoparticle  occur in the $XOZ$ plane (b). We assume that the distance between magnetic nanoparticles is large enough, so we neglect the interaction between magnetic nanoparticles. Schematics of double-well potential in a ferroelectric substrate (c).}
\label{fig:schematic plot}
\end{figure*}

The pictorial plot of the system of interest is shown in Fig.\ref{fig:schematic plot}. 
ME coupling between two nonlinear magnetic and ferroelectric subsystems promises nontrivial dynamical effects.  The study of those effects is in the scope of the present work. Our model is inherently nonlinear and we analyze it in different limits of weak, strong, and moderate nonlinearity. Despite the significant interest in ME coupling, dynamic aspects of the coupled ferroelectric crystal-magnetic nanoparticle system still need to be studied in general and rigorous mathematical form. Present work aims to fill this gap. The work is organized as follows: Section \textbf{II} specifies the model. We show that by applying two time-dependent electric fields with different frequencies, we can dynamically design the confinement potential of the ferroelectric subsystem and change its shape from quintic to double-well potential. It allows us to realize two different types of dynamics. In particular, we show that the system is more sensitive concerning the ME coupling in the case of the double-well potential.  
Section \textbf{III} considers the Hamiltonian approach and the case of moderate and strong nonlinearity. In particular, we neglect dissipation processes in both the ferroelectric and magnetic parts. This approximation is valid for relatively short-time dynamics and, for example, for Yttrium iron garnet, a material characterized by small Gilbert damping. Besides, through the work, we neglect the factor of thermal noise, and therefore our discussion is adapted to the zero and low-temperature limit cases. We implement the Kolmogorov Arnold Moser theory tools, exploit canonical action-angle variables, and study the overlapping of nonlinear resonances. We find that the hybrid ferroelectric-ferromagnetic system is characterized by two invariant tors (one tori per subsystem). Switching on the ME coupling destroys the tors. Analyses of Melnikov's function show that in the limit of moderate nonlinearity, the dynamics of the magnetic subsystem is chaotic, while the dynamics of the ferroelectric subsystem is regular. In the case of strong nonlinearity, analysis of the phase space region near the separatrix shows the formation of a homoclinic structure, and the dynamics of both ferroelectric and magnetic subsystems are chaotic even for an arbitrary small ME coupling term. 
In section \textbf{IV}, we study bifurcations in the system. Analysis of the phase space of the hybrid system shows that Hopf bifurcations occur in the system when tuning the amplitude of the ME coupling term.  
In section \textbf{V}, we address the problem of weak nonlinearity. We assume that the deviation of the system from equilibrium is relatively small and implement Van der Pol's method in the non-resonant case. In contrast, in the resonant case, we implement Bogoliubov's method. We linearize the system near a particular nonlinear resonance and explore the overlapping of the nonlinear resonances. We find that the parametric resonance problem describes the magnetic subsystem's dynamic. We find conditions when oscillations in the ferroelectric subsystem parametrically enhance oscillations in the magnetic subsystem. Solving the Mathieu equation, we find the external field's frequency when the hybrid system's dynamic is periodic. In section \textbf{VI}, we analyze the problem of strong nonlinearity and dissipation processes in the system. We study Lyapunov's function and fractal dimension and find a strong dependence of both quantities on the ME coupling term. 

\section{Model}
The free energy of the entire system consists of three parts: The ferroelectric part $F_f$, magnetic part $F_m$, and the ME coupling term $V$:
\begin{eqnarray}\label{we specify below}
F=F_f+F_m+V_{ME}.
\end{eqnarray}
The free energy of the magnetic nanoparticle has the form $F_m=\frac{1}{2}Km_y^2-h_zm_z$, where $K$ is the anisotropy constant, $\textbf{m}=\textbf{M}/\vert\textbf{M}\vert$ is the unit magnetization vector, $\textbf{h}=(0,0,h_z)$ is the external magnetic field. The free energy of the ferroelectric part has the form $F_f=\frac{\dot{x}^2}{2}-\frac{\alpha}{2}x^2+\frac{\beta}{4}x^4-xE(t)$. Taking into account ME coupling term $V_{ME}=-g_{ME}xm_z$, 
dynamical equations governed by the free energy Eq.(\ref{we specify below}) read:
\begin{eqnarray}\label{equations before}
&& \frac{d\textbf{m}}{dt}=-\textbf{m}\times \textbf{h}_{eff}+\alpha_G\left(\textbf{m}\times \frac{d\textbf{m}}{dt}\right),\nonumber\\
&& \frac{d^2x}{dt^2}=-\frac{\delta F}{\delta x}-\frac{\delta V_f}{\delta x},\\
&& \textbf{h}_{eff}=-\frac{\delta F}{\delta \textbf{m}}.\nonumber
\end{eqnarray}
Here $\alpha_{G}$ is the Gilbert damping constant, $-\frac{\delta V_f}{\delta x}=V_0\sin(\omega t)+C\sin(\Omega t)$ are two periodic external driving fields applied to the ferroelectric subsystem. In what follows, we adopt the ansatz \cite{gibson2020nonlinear}:
$m=\left(\cos\theta\sin\varphi,\sin\theta,\cos\theta\cos\varphi\right)$, $-\pi/2\leqslant\theta\leqslant\pi/2$, $-\pi<\varphi\leqslant\pi$. Further, we assume that
$h_s/K\ll 1$, $h_d/K\ll 1$  and therefore $m_y\approx 0$, $\theta\approx 0$. Then from Eq.(\ref{equations before}) we deduce the corresponding Lagrangian and Hamiltonian functions:
\begin{eqnarray}\label{Corresponding Lagrangian and Hamiltonian}
&&\mathcal{L}=\frac{1}{2}\dot{\varphi}^2+\frac{1}{2}\dot{x}^2+\omega_0^2\cos\varphi-\frac{\alpha}{2}x^2+\frac{\beta}{4}x^4+\nonumber\\
&&g_{ME}x\cos\varphi+xV_0\sin\omega t+xC\sin\Omega t,\nonumber\\
&&H\left(\lbrace p_n\rbrace, \lbrace q_n\rbrace\right)=\sum\limits_np_n\dot{q}_n-\mathcal{L},\,\,\,q_n=\lbrace \varphi,x\rbrace.
\end{eqnarray}
Here $\omega$ and $\Omega$ are the frequencies of the driving external electric fields.
The Lagrangian equations of motion have the form:
\begin{eqnarray}\label{coupled}
&&\frac{d}{dt}\frac{\partial \mathcal{L}}{\partial \dot{q}_n}=\frac{\partial\mathcal{L}}{\partial q_n}-\frac{\partial D}{\partial \dot{q}_n},
\end{eqnarray}
where $D=\frac{\alpha_G\dot{x}^2+\alpha_G\dot{\varphi}^2}{2}$
is the dissipative function. In what follows, our primary focus lies in Eq.~(\ref{coupled}). We will analyze Eq.(\ref{coupled}) 
in physically relevant different cases. 
Following \cite{gitterman2001bistable,landa2000vibrational}, we assume that $C>\alpha/4\beta$ and $\Omega\gg\omega$ and look for the solution of the ferroelectric subsystem in the following form:
\begin{eqnarray}\label{ansatz}
x(t)=y(t)-\frac{C\sin(\Omega t)}{\Omega^2}.
\end{eqnarray}
After substituting Eq.(\ref{ansatz}) into Eq.(\ref{coupled}) and performing the averaging over the high frequency, 
for a averaged variable $X(t)=\langle y(t)\rangle$, for the case $E=0$  we obtain an effective equation
\begin{eqnarray}\label{an effective equation}
&&\frac{d^2X}{dt^2}+\alpha_G\frac{dX}{dt}+\gamma X+\beta X^3-g_{ME}\cos\varphi=V_0\sin(\omega t),\nonumber\\
&&\frac{d^2\varphi}{d\tau^2}+\alpha_G\frac{d\varphi}{d\tau}=-\sin\varphi-g_{ME}X\sin\varphi.
\end{eqnarray}
Here $\gamma=\left(\frac{3\beta C^2}{2\Omega^4}-\alpha\right)$ is a rescaled linear frequency, and in the ME coupling term, we considered coupling to the averaged variable $X$ instead of fast oscillating $x$. We added the damping term to the ferroelectric subsystem and assumed that the damping rate is comparable to the magnetic part. If the magnetic nanoparticle is the essence of the small yttrium-iron-garnet (YIG) sphere, the damping constant is too small and can be neglected for a short time dynamics \cite{zhang2015cavity}.

\section{Hamiltonian approach}

After converting the initial problem into the problem of nonlinear oscillator Eq.(\ref{an effective equation}), we apply the methods of nonlinear resonance and KAM theory \cite{arnol2013mathematical,PhysRevB.101.104311,singh2022hybrid}. We assume that damping in the system is small and can be neglected. On the other hand, we assume that nonlinearity in the system is strong and nonlinear terms cannot be tackled  perturbatively. Keeping in mind that $\gamma=\left(\frac{3\beta C^2}{2\Omega^4}-\alpha\right)$ at first we consider the case $\gamma>0$. \vspace{0.3cm}\\
\textbf{Moderate nonlinearity} \vspace{0.2cm}\\
We consider the problem of nonlinear resonance and dynamics of the system in the vicinity of particular resonance. Results obtained in this paragraph are valid in the limit of moderate nonlinearity. The condition of moderate nonlinearity is quantified below.  
We consider the system's dynamics in the vicinity of the particular nonlinear $n$th resonance in the system and apply the non-perturbative method valid in the case of moderate nonlinearity \cite{PhysRevB.101.104311}.
For convenience, we switch to the canonical pair of  action-angle $(I, \theta)$ variables. The cantilever part of the Hamiltonian $H_{p,q}=H_{0}+H_{NL}+V(x,t)$ expressed in new variables $H_{I,\theta}$ is connected to the original Hamiltonian through the production function $\Phi=F+I\theta$ via the relation:
\begin{eqnarray}\label{L3}
&& d\Phi=pdq+\theta dI+\big(H_{I,\theta}-H_{p,q}\big)dt,
\end{eqnarray}
and the canonical set of equations in new variables are:
\begin{equation}\label{L4}
  \left.\begin{aligned}
  \frac{dI}{dt}&=-\frac{\partial H_{I,\theta}}{\partial\theta}=- \frac{\partial V(I,\theta, \lambda)}{\partial \theta}, \\
  \frac{d\theta}{dt}&=\frac{\partial H_{I,\theta}}{\partial I}=\Theta(I)+ \frac{\partial V(I,\theta, \lambda)}{\partial I}.
\end{aligned}\right\} 
\end{equation}
Here we introduced the nonlinear frequency $\Theta(I)=\partial (H_{0}+H_{NL})/\partial I$, $\dot{\lambda}=\omega$ the frequency of external driving and transformed Hamiltonian is defined as follows $H_{I, \theta}=H_0+H_{NL}$,
$H_0=\sqrt{2\gamma}I$, $H_{NL}=3\pi\beta\left(I^2/2\gamma\right)^2$, $V=V_0\sqrt{I/\sqrt{2\gamma}}$. An important fact is that nonlinear frequency of oscillation $\Theta(I)$
is a function of the action variable $I$. We introduce characteristic measure of nonlinearity \cite{zaslavsky2007physics}
$\mathcal{M}=\left|\frac{I}{\Theta}\frac{d\Theta}{dI}\right|$. The results we are going to present in this paragraph are valid for $V_0\ll \mathcal{M}\ll 1/V_0$. The set of equations Eq.(\ref{L4}) is governed by the effective Hamiltonian $H_{eff}$. We skip the details of the cumbersome calculations \cite{PhysRevB.101.104311} and write the final result:
\begin{eqnarray}\label{the effective Hamiltonian}
H_{eff}=\varrho(\Delta I)^2/2+V(I_n)\cos\psi_n.  
\end{eqnarray}
Here $\Delta I=I-I_n$ is the deviation of the action variable from the $n$th nonlinear resonance, $\varrho=(\frac{d\theta(I)}{dI})_{I=I_n}$ and 
$\psi_n=\theta-n\omega t$. For the sake of brevity in what follows we omit the index $n$. 
The system of our interest, single crystal TGS coupled to the YIG sphere we recast in the forms:
\begin{eqnarray}\label{Hamiltonian case first}
&&\frac{d^2\varphi}{d\tau^2}+\left(1-g'_{ME}\Delta I\right)\sin\varphi=0,\nonumber\\
&&\Delta \dot{I}=V\sin\psi-g_{ME}\cos\varphi,\nonumber\\
&& \dot{\psi}=\varrho\Delta I.
\end{eqnarray}
Here we averaged over the fast oscillating phase of the variable $X$ and took RMS amplitude of $X$ in the ME coupling term  $g'_{ME}=-g_{ME}/\sqrt{\gamma}$.  
We analyze Eq.(\ref{Hamiltonian case first}) iteratively. At first we solve Eq.(\ref{Hamiltonian case first}) in the limit $g_{ME}\rightarrow 0$ and for a ferroelectric subsystem we obtain: 
\begin{widetext}
\begin{eqnarray}\label{Hamiltonian solution ferro}
&&\Delta I_+(t,k_f)=\sqrt{(E_f+V)/\varrho}\,\,\text{dn}[\varrho\sqrt{(E_f+V)/\varrho}\,\,t,k_f],\,\,\,\,\,\,\,E_f>V,\nonumber\\
&&\Delta I_-(t,k_f)=\sqrt{(E_f+V)/\varrho}\,\,\text{cn}[\varrho\sqrt{(E_f+V)/\varrho}\,\,t,1/k_f],\,\,E_f<V,\nonumber\\
&&\psi_<(t,k_f)=\frac{1}{k_f}\arccos[\text{dn}(t,k_f)],\,\,\,k_f<1,\,\,\,\,t=t\sqrt{V\varrho},\nonumber\\
&&\psi_>(t,k_f)=\arcsin[\text{sn}(t,1/k_f)],\,\,\,k_f>1.
\end{eqnarray}
\end{widetext}
Here $\text{dn(...)}$, $\text{cn(...)}$ are Jacobi elliptic functions, $E_f$ is the energy of the ferroelectric subsystem, $k_f=\sqrt{2V/(E_f+V)}$ is the parameter, $\Delta I_+$, $\Delta I_-$ are two topologically different solutions divided from each other by separatrix $E_f\rightarrow V,\,\,k\rightarrow 1$.
Similarly for a magnetic subsystem, we deduce;
\begin{eqnarray}\label{Hamiltonian solution magnetic}
&&\dot{\varphi}_<(t,k_m)=2k_m \text{cn}(t,k_m),\,\,\,k_m<1\nonumber\\
&&\dot{\varphi}_>(t,k_m)=2k_m \text{dn}(t,1/k_m),\,\,\,k_m>1,\\
&&\varphi_<(t,k_m)=2\arccos[\text{dn}(t,k_m)],\,\,\,k_m<1,\nonumber\\
&&\varphi_>(t,k_m)=2k_m\arcsin[\text{sn}(t,1/k_m)],\,\,\,k_m>1.\nonumber
\end{eqnarray}
Here $k^2_m=\frac{1}{2}(1+E_m)$ and $E_m$ is the energy of the magnetic subsystem. 
Existence of the two separatrix lines in Eq.(\ref{Hamiltonian solution ferro}), Eq.(\ref{Hamiltonian solution magnetic}) hints that ME perturbation term will lead to the chaos. To explore the criteria of chaos, we apply 
We utilize the Melnikov function to determine a measure of distance between stable and unstable manifolds in the Poincaré map in ferroelectric and magnetic subsystems, respectively. The Melnikov function for our problem can be evaluated under a particular approximation. Namely, we evaluate the Melnikov function for ferroelectric and magnetic systems separately. When evaluating the Melnikov function for the ferroelectric (magnetic) system, in the ME coupling term $V_{ME}=-g_{ME}X\cos\varphi$ we consider unperturbed $g_{ME}=0$ solution of the magnetic Eq.(\ref{Hamiltonian solution magnetic}) (ferroelectric Eq.(\ref{Hamiltonian solution ferro})) subsystem as an external time-dependent perturbation. Consequently, we deduce.
\begin{eqnarray}\label{Melnikov function}
&& D_f=-g_{ME}\varrho\int\limits_{-\infty}^{\infty}\Delta I_{<,(>,)}(t)\cos\varphi_{<,(>)}(t)dt,\\
&&D_M=g'_{ME}\int\limits_{-\infty}^{\infty}\Delta I_{<,(>,)}(t)\dot{\varphi}_{<,(>)}(t)\sin\varphi_{<,(>)}(t)dt.\nonumber
\end{eqnarray}
As we see from Eq.(\ref{Melnikov function}) Melnikov function estimated for the ferroelectric system is directly proportional to the ME coupling, and ME coupling is a source of chaos in the system. 
To evaluate the Melnikov function in the vicinity of the separatrix, we consider asymptotic $k_f,k_m\rightarrow 1$
and shift of the time $t\rightarrow t+\tau$. Then from Eq.(\ref{Melnikov function}) we deduce
\begin{eqnarray}\label{Melnikov intermediate}
&&\frac{-D_f(\tau)}{g_{ME}\sqrt{V\varrho}}\approx \int\limits_{-\infty}^{\infty}\frac{\cos[2\arccos(1/\cosh(t))]}{\cosh[\sqrt{2V\varrho}(t+\tau)]}dt,\\
&&\frac{D_D(\tau)}{2g'_{ME}\sqrt{V/\varrho}}\approx \int\limits_{-\infty}^{\infty}\frac{\sin[2\arccos(1/\cosh(t+\tau))]}{\cosh[\sqrt{2V\varrho}t]\cosh[t+\tau]}dt.\nonumber
\end{eqnarray}

Here $\cosh(...)$ is the hyperbolic function. 
After performing integration in Eq.(\ref{Melnikov intermediate}) finally we obtain:
\begin{eqnarray}\label{Melnikov final}
&& D_f(\tau)\approx 4g_{ME}\pi\left(\frac{1}{2}+\frac{1-\text{cosh}(\tau)}{\text{sinh}^2(\tau)}\right),\\
&&D_D(\tau)\approx 8\pi g'_{ME}\sqrt{\frac{V}{\varrho}}\left(\frac{4\text{cosh}(\tau)-3-\text{cosh}(2\tau)}{4\text{sinh}^3(\tau)}\right).\nonumber
\end{eqnarray}
The emergence of chaos can be identified as the change in the sign of the
Melnikov function in Eq.(\ref{Melnikov final}). It is easy to see that $D_f(\tau)$ in Eq.(\ref{Melnikov final}) is always positive, while $D_m(\tau)$ changes the sign for $\tau=0$. Thus, in the limit of moderate nonlinearity, the ferroelectric subsystem dynamics are more stable than the magnetic nanoparticle. \vspace{0.3cm}\\
\textbf{The strong nonlinearity case}\vspace{0.3cm}\\
We proceed with analyzing the case $\gamma<0$ in the limit of strong nonlinearity. The equation of the unperturbed ferroelectric subsystem (hereafter $|\gamma|=\beta=1$):

\begin{eqnarray}\label{unperturbed ferroelectric subsystem}
&&\ddot{X}=|\gamma| X-X^3,
\end{eqnarray}
has a saddle fixed point $(X,\dot{X})=(0,0)$ and the center $(\pm 1,0)$. The saddle point is characterized by homoclinic orbits
\begin{eqnarray}
X_s^{\pm}(t)=\pm (\sqrt{2}\text{sech}(t),-\sqrt{2}\text{sech}(t)\text{tanh}(t)).
\end{eqnarray}
Inside the region of the homoclinic orbits, the solution reads
\begin{equation}\label{Inside homoclinic orbits}
X_{\pm}=\left\{\begin{aligned}
  \frac{\pm\sqrt{2}}{\sqrt{2-k^2_f}}\text{dn}\left(\frac{t}{\sqrt{2-k_f^2}},k_f\right),\,\,\,\,\,\,\,\,\,\,\,\,\,\,\,\,\,\,\,\,\,\,\,\,\,\,\,\,\\
  \frac{\mp\sqrt{2}k^2_f}{2-k^2_f}\text{sn}\left(\frac{t}{\sqrt{2-k^2_f}}\right)\text{cn}\left(\frac{t}{\sqrt{2-k^2_f}}\right).
\end{aligned}
\right.
\end{equation}
Here $\pm$ corresponds to two different centers $(\pm 1,0)$. The period of oscillation is given by $T(k_f)=2K(k_f)\sqrt{2-k^2}$, where $K(k_f)$ is the complete elliptic of the first kind, $E_f=\frac{k^2_f-1}{(2-k^2)^2}$ is the energy of the ferroelectric subsystem. To find criteria for the zeros of the Melnikov function for the ferroelectric subsystem, linearized solution of the magnetic subsystem $\varphi(t)=\varphi_0\sin(t),\,t<1/\alpha_G$ we consider as a time-dependent perturbation for the ferroelectric part. Following \cite{greenspan1984repeated} for Melnikov's zeros, we deduce:
\begin{eqnarray}\label{ferroelectric Melnikov's zeros}
\left\vert\frac{4\alpha_G\left[(2-k^2)E(k)-2(1-k^2)K(k)\right]}{3g_{ME}\sqrt{2}\pi(2-k^2)^{3/2}\text{Sech}\frac{\pi mK(\sqrt{1-k^2})}{K(k)}}\right\vert<1,
\end{eqnarray}
where $\pi m=K(k)\sqrt{2-k^2}$.
Eq.(\ref{ferroelectric Melnikov's zeros}) defines minimal values of the $g_{ME}$ for chaos. In the vicinity of the separatrix
$g_{ME}>\frac{2\sqrt{2}\alpha_G}{\pi}\text{cosh}(\pi/2)$. Taking into account the small value of the Gilbert damping {\it e. g.}, for YIG $\alpha_G=0.027$ \cite{yamada2020dependence}, we see that the chaos sets in the system even for small $g_{ME}=0.06$. 

\section{ Hopf bifurcation}
In this section, we study in detail the phase portrait of the system and analyze the Hopf bifurcation caused by the ME coupling term. In the absence of the ME coupling term $g_{ME}$, and dissipation phase portrait of the magnetic nanoparticle is quite simple \cite{zaslavsky2007physics}. The analysis of eigenvalues of the Jacobian shows that at the fixed point $(\varphi,\,\dot{\varphi})=(0,0)$, the determinant of the matrix is one. The determinant greater than zero means that the fixed point is either a center or a spiral. On the other hand, it is easy to see that the sum of eigenvalues $\lambda_1+\lambda_2=0$ (i.e., the fixed point is a stable center).

\begin{figure}[!ht]
\centering
\includegraphics[width=\columnwidth]{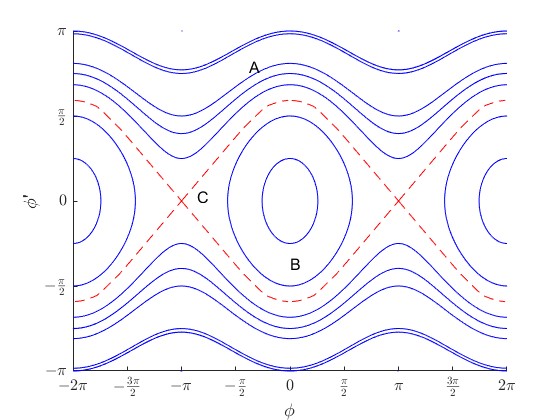}
\caption{The phase portrait of the magnetic subsystem in the absence of the dissipation and magnetoelectric coupling. We see two topologically distinct types of trajectories: $\mathcal{A}$ open trajectories are separated from the closed trajectories $\mathcal{B}$ by the separatrix line $\mathbb{C}$.}
\label{fig:phase portrait no damping}
\end{figure}

\begin{figure}[!ht]
\centering
\includegraphics[width=\columnwidth]{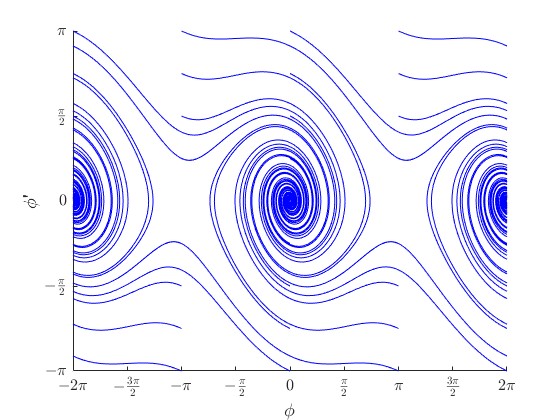}
\caption{The phase portrait of the magnetic subsystem, the effect of the dissipation term. We clearly see that the fixed point is a stable spiral.}
\label{fig:phase portrait with damping}
\end{figure}

For the fixed point $(0,\pi)$, the determinant of the Jacobian matrix is less than zero $\text{det}(J)<0$, meaning that we have a saddle. Phase portrait of the magnetic nanoparticle is the essence of the heteroclinic orbits  $\mathbb{C}$ connecting two saddle points Fig.\ref{fig:phase portrait no damping}. Heteroclinic orbits  $\mathbb{C}$ split the phase space in two topologically distinct phase trajectories open $\mathcal{A}$ and closed $\mathcal{B}$ trajectories. The Gilbert damping modifies the phase portrait of the magnetic nanoparticle. The effect of the dissipation term is plotted in Fig. \ref{fig:phase portrait with damping}. We see that the fixed point is a stable spiral.

We aim to explore the impact of the ME coupling term on the phase portrait of the system. For this purpose, we analyze the system of equations:

\begin{eqnarray}\label{Hopf begining}
&&\frac{d^2\varphi}{d t^2}+\alpha_G\dot{\varphi}+\left(1-g'_{ME}\Delta I\right)\sin\varphi=0,\nonumber\\
&&\Delta \dot{I}=V\sin\psi-g_{ME}\cos\varphi,\nonumber\\
&& \dot{\psi}=\varrho\Delta I.
\end{eqnarray}
In the limit $V\gg G_{ME}$ we can solve Eq.(\ref{Hopf begining}) iteratively. Taking into account
Eq.(\ref{Hamiltonian solution ferro}), we estimate the maximal value of the modulation of the adiabatic variable 
$\langle\Delta I\rangle$

\begin{eqnarray}\label{adiabatic variable}
&&\langle\Delta I\rangle=\frac{\sqrt{(E_f+V)/\rho}}{K(k)}\int\limits_0^{4K(k)}\text{dn}(t,k)dt=\nonumber\\
&&\frac{\pi\sqrt{(E_f+V)/\rho}}{2K(k)}.
\end{eqnarray}
Here $K(k)$ is the second order complete elliptic integral. Taking into account Eq.(\ref{adiabatic variable})  the Jacobian matrix of the magnetic nanoparticle takes the form
\begin{eqnarray}\label{Jacobian}
\begin{bmatrix}
-1 & -(1-g'_{ME}\langle\Delta I\rangle)\cos\varphi \\
1 & 0
\end{bmatrix}.
\end{eqnarray}
For \label{Jacobian} we study four cases of particular interest:\vspace{0.5cm}\\
\textbf{[A]}. Fixed point $(0,0)$,  $g'_{ME}=0$,  $\text{det}(J)=1>0$. Because of
$\lambda_1=-\frac{1}{2}+\frac{\sqrt{3}}{2}i$, $\lambda_2=-\frac{1}{2}-\frac{\sqrt{3}}{2}i$, and
$\text{Re}(\lambda_1+\lambda_2)=-1<0$ the fixed point is a stable spiral. \vspace{0.5cm}\\
\textbf{[B]}. Fixed point $(0,\pi)$,  $g'_{ME}=0$,  $\text{det}(J)=-1<0$. Because of
$\lambda_1=-\frac{1}{2}-\frac{\sqrt{5}}{2}$, $\lambda_2=-\frac{1}{2}+\frac{\sqrt{5}}{2}$, and
$\text{Re}(\lambda_1+\lambda_2)=-1<0$ the fixed point is a saddle.\vspace{0.5cm}\\
\textbf{[C]}. Fixed point $(0,0)$,  $g'_{ME}\neq0$,  $\text{det}(J)=1-g'_{ME}\langle\Delta I\rangle$. Because of $\lambda_1=\frac{1}{2}\left(-\sqrt{4g'_{ME}\langle\Delta I\rangle-3}-1\right)$, $\lambda_2=\frac{1}{2}\left(\sqrt{4g'_{ME}\langle\Delta I\rangle-3}-1\right)$. For $g'_{ME}\langle\Delta I\rangle<1$ the fixed point is a stable spiral, $g'_{ME}\langle\Delta I\rangle>1$ the fixed point is a saddle.\vspace{0.5cm}\\
\textbf{[D]}. Fixed point $(0,\pi)$,  $g'_{ME}\neq0$,  $\text{det}(J)=g'_{ME}\langle\Delta I\rangle-1$. Because of $\lambda_1=\frac{1}{2}\left(\sqrt{5-4g'_{ME}\langle\Delta I\rangle}-1\right)$, $\lambda_2=\frac{1}{2}\left(-\sqrt{5-4g'_{ME}\langle\Delta I\rangle}-1\right)$. For $g'_{ME}\langle\Delta I\rangle<1$ the fixed point is a stable spiral, $g'_{ME}\langle\Delta I\rangle>1$ the fixed point is a saddle.\vspace{0.5cm}\\
\textbf{Bifurcations in the system}:\vspace{0.5cm}\\
\textbf{[i]}. Depending on the value of $g'_{ME}\langle\Delta I\rangle$ we have a transition from the 
stable spiral fixed point $g'_{ME}\langle\Delta I\rangle<1$  into the saddle fixed point $g'_{ME}\langle\Delta I\rangle>1$. \vspace{0.5cm}\\
\textbf{[ii]}. If the fixed point is stable, the eigenvalues must both lie in the left half-plane 
$\text{Re}(\lambda_1)<0$, $\text{Re}(\lambda_2)<0$. Thus Hopf bifurcation in the system occurs when 
$5-4g'_{ME}\langle\Delta I\rangle<1$ changes to $5-4g'_{ME}\langle\Delta I\rangle>1$ and 
$4g'_{ME}\langle\Delta I\rangle-3<1$ to $4g'_{ME}\langle\Delta I\rangle-3>1$. Thus we conclude that the system is pretty mach sensitive to the ME interaction.\vspace{0.5cm}\\
\textbf{KAM theorem and magneto-electric coupling:}\vspace{0.5cm}\\
In the spirit of the KAM theorem, we neglect the dissipation in the system and present the total Hamiltonian in terms of the canonical 
action-angle variables:
\begin{eqnarray}\label{KAM1}
&&H_{tot}=H_0+U,\nonumber\\
&&H_0=\frac{\rho(\Delta I)^2}{2}+V\cos\psi+\frac{1}{2}\dot{\varphi}^2-\cos\varphi,\nonumber\\
&&U=-g_{ME}\Delta I\cos\varphi.
\end{eqnarray}
Here in Eq.(\ref{KAM1}) $(\Delta I, \psi)$ are the canonical action-angle variables of the ferroelectric subsystem, while the canonical conjugate 
of the magnetic angular variable $\varphi$ has a form $I(E_m)=\frac{2}{\pi}\int\limits_0^{\varphi_0}d\varphi\left[ 2(E_m+\cos\varphi)\right]^{1/2} $, $\varphi_0=\text{arccos}(-E_m)$ 
and can be calculated through the complete elliptic integrals by following formulae 
\begin{eqnarray}\label{KAM2}
&&I=\frac{8}{\pi}E\left(\frac{\pi}{2},k_m\right)-(1-k_m^2)F\left(\frac{\pi}{2},k_m\right),\,\,\,k_m<1\nonumber\\
&&I_m=\frac{8}{\pi}k_mE\left(\frac{\pi}{2},\frac{1}{k_m}\right),\,\,k_m>1,
\end{eqnarray}
where $k_m^2=(1+E_m)/2$ and $E_m$ is the energy of the magnetic subsystem. To explore degeneracy in the system, we calculate the following Hessian 
\begin{eqnarray}\label{KAM3}
\mathcal{D}=\text{det}\begin{bmatrix}
\frac{\partial^2 H_0}{(\partial\Delta I)^2} & \frac{\partial^2H_0}{\partial\Delta I\partial I}\\
\frac{\partial^2H_0}{\partial I\partial\Delta I} & \frac{\partial^2 H_0}{\partial T^2}
\end{bmatrix}.
\end{eqnarray}
To calculate $\frac{\partial^2 H_{0}}{\partial I^2}$ we introduce the nonlinear frequency of the magnetic subsystem:
\begin{eqnarray}\label{nonlinear frequency}
  \omega_m (H_0)&=&\frac{d(H_0)}{dI}=\left[ \frac{d I}{dH_{0}}\right]^{-1},\nonumber  \\
  \omega_m (H_0) &=& \frac{\pi}{2}\frac{1}{F(\frac{\pi}{2},k_{m})},\,\,\,\,k_{m}<1,\\
   \omega_m (H_0) &=& \frac{\pi}{2}\frac{k_m}{F(\frac{\pi}{2},\frac{1}{k_m})},\,\,\,\,k_{m}>1, \nonumber
\end{eqnarray}
where $F(\frac{\pi}{2},k)$ is the complete elliptic integral of first order.
Taking into account Eq.(\ref{nonlinear frequency}) we obtain:
\begin{widetext}
\begin{eqnarray}\label{2}
\mathcal{D}=\left\{  \begin{array}{ll}
\frac{\rho}{2k_m^2}\omega_m^2(H_0)\big(1-\frac{2}{\pi}\frac{E(\frac{\pi}{2},k_m)}{(1-k_m^2)}\omega_m(H_0)\big),\,\,k_m<1,\\
\frac{\rho}{4k_m^2}\omega_m^2(H_0)\bigg(1+\frac{1}{k_m^2}\big(\frac{2}{\pi}k_{m}\omega_m(H_0)\frac{E(\frac{\pi}{2},\frac{1}{k_m})}{(1-\frac{1}{k_m^2})}-1\big)\bigg),\,\,k_m>1.
\end{array}
\right .
\end{eqnarray}
\end{widetext}
Here $E(\frac{\pi}{2},k)$ is the complete elliptic integrals of second order.
For further analysis of Eq.(\ref{2}) we exploit the asymptotic of the elliptic functions for large and small argument $k_m$ and deduce
\begin{eqnarray}\label{KAM5}
&&\mathcal{D}(k_m\ll 1))=-\frac{3\rho}{8},\nonumber\\
&&\mathcal{D}(k_m\gg 1))=\frac{\rho}{2},\nonumber\\
&&\mathcal{D}(k_k=1)=0.
\end{eqnarray}
The obtained result Eq.(\ref{KAM5}) means that in the vicinity of the separatrix $k_m=1$, most of the invariant tors are destroyed and strong chaos sets in the system.

\section{Weak nonlinearity}
This section analyzes the system in the limit of weak nonlinearity when the system's dynamic is regular. We consider dynamics close to the equilibrium point $\varphi\ll 1$  adopt the ansatz $\cos\varphi=1-\varphi^2/2$,  $\sin\varphi=\varphi$. We assume that $\gamma|X|>\beta|X|^3$ and rewrite the system of equations in the following form
\begin{eqnarray}\label{Butenin case}
&&\frac{d^2X}{dt^2}+\alpha_G\frac{dX}{dt}+\gamma X+\beta X^3-\nonumber\\
&& g_{ME}(1-\varphi^2/2)=V_0\sin(\omega t),\nonumber\\
&&\frac{d^2\varphi}{d t^2}+\alpha_G\frac{d\varphi}{d t}=-\varphi+g_{ME}X \varphi.
\end{eqnarray} 
In what follows, we exploit Van der Pol's method generalized for the systems with several degrees of freedom  \cite{chotorlishvili2011nonlinear}.
We rewrite Eq.(\ref{Butenin case}) in the form
\begin{eqnarray}\label{rewrite Butenin}
&&\frac{d^2X}{dt^2}+\gamma X= N(X,\dot{X},\varphi, \dot{\varphi},t),\nonumber\\
&&\frac{d^2\varphi}{d t^2}+\varphi= M(X,\dot{X},\varphi, \dot{\varphi},t),
\end{eqnarray} 
where $N(X,\dot{X},\varphi, \dot{\varphi},t)$ and $M(X,\dot{X},\varphi, \dot{\varphi},t)$ are nonlinear parts considered as small corrections to the linear equations. At first, we consider nonresonant case $\sqrt{\gamma}\neq\omega\neq 1$ and look for the solution of the linear part in the following form:
$\varphi_0(t)=A_1\sin(t+\alpha_1)$, $X_0=A_2\sin(\sqrt{\gamma}t+\alpha_2)+B\sin\omega\tau$, $B=\frac{V_0}{\gamma-\omega}$. After inserting the solutions of the linear part into the Eq.(\ref{rewrite Butenin}) we deduce equations for the amplitudes and phases
\begin{eqnarray}\label{amplitude and phase Butenin}
&& \dot{A}_1(t)=- N(X,\dot{X}_0,\varphi_0, \dot{\varphi}_0,t)\cos(t+\alpha_1),\nonumber\\
&&\dot{\alpha}_1(t)A_1(t)= N(X,\dot{X}_0,\varphi_0, \dot{\varphi}_0,t)\sin(t+\alpha_1),\nonumber\\
&&\dot{A}_2=-\frac{1}{\sqrt{\gamma}}M(X,\dot{X}_0,\varphi_0, \dot{\varphi}_0,t)\cos(\sqrt{\gamma}t+\alpha_1),\\
&&\dot{\alpha}_2(t)A_2(t)=\frac{1}{\sqrt{\gamma}}M(X,\dot{X}_0,\varphi_0, \dot{\varphi}_0,t)\sin(\sqrt{\gamma}t+\alpha_1).\nonumber
\end{eqnarray}
We average Eq.(\ref{amplitude and phase Butenin}) over the fast phases. We skip the technical details and present the final result:
\begin{eqnarray}\label{solution amplitude and phase Butenin}
&& \varphi(t)=A_1(0)e^{-\frac{\alpha_gt}{2}}\sin(t+\alpha_1(0)),\nonumber\\
&& X(t)=A_2(0)e^{-\alpha_gt/2}\sin\bigg[\left(\sqrt{\gamma}-\frac{3\beta}{4\sqrt{\gamma}}B^2\right)t-\nonumber\\
&&\frac{\beta}{4\sqrt{\gamma}\alpha_g}A_2^3(0)\left(1-e^{-3\alpha_gt/2}\right)+\alpha_2(0)\bigg]+\nonumber\\
&&B\sin\omega t. 
\end{eqnarray}
From Eq.(\ref{solution amplitude and phase Butenin}), we see that in the case of a weak nonlinearity, the phase trajectory of the magnetic subsystem is a spiral sink, while for the ferroelectric subsystem, we have a limit cycle. For more insights into the problem, to analyze the nonlinear resonance problem we apply Bogoliubov's method \cite{bogoliubov1961asymptotic} and solve the problem iteratively, looking for the higher iteration terms. In particular, we consider solutions to the linear equations in the ME coupling term $V_{ME}(X_0,\varphi_0)$. We assume that the frequencies of the magnetic and ferroelectric subsystems are commensurate and, therefore, in the linear approximation, the ME term is a periodic function with the period $T=2\pi/\omega_{mf}$. 
We tackle the ferroelectric subsystem in the following form \cite{bogoliubov1961asymptotic}:
\begin{eqnarray}\label{Bogoliubov1}
&&\ddot{X}+\omega_0^2X=\beta Q\left(X,\dot{X},\omega t\right),
\end{eqnarray}
where $Q\left(X,\dot{X},\omega t\right)=-g_{ME}/\beta-X^3-(\alpha_G/\beta)\dot{X}+(V_0/\beta)\sin\omega t$, and $\omega_0^2=\gamma$. 
We assume that $\omega_0^2=\left(\frac{p}{q}\,\omega\right)^2+\beta\Delta$, where $p,q$ are integers and $\Delta$ is a small detuning. 
In Eq.(\ref{Bogoliubov2}) the unknown amplitude and the phase of nonlinear oscillations are defined from the set of equations
\begin{eqnarray}\label{Bogoliubov3}
&&\dot{a}=\beta f_1(a,\theta),\nonumber\\
&&\dot{\psi}=\frac{p}{q}\omega_0+\beta\omega_1(a,\theta),
\end{eqnarray}
where $\psi=\frac{p}{q}\omega t+\theta$.
We will look for the solution of Eq.(\ref{Bogoliubov1}) in the following form:
\begin{eqnarray}\label{Bogoliubov2}
X(t)=a\cos\psi+\beta X_1(a,\theta,\omega t)\ldots
\end{eqnarray}
The first order perturbation term $\beta Q_0(a\cos\psi,-\omega_0 a\sin\psi, \omega)$ we expand into the double Fourier series:
\begin{eqnarray}\label{Bogoliubov4}
&&\beta Q_0(a,\psi,\omega t)=\beta\sum\limits_{n,m}\alpha_{nm}(a)e^{i(n\omega t+m\psi)},\nonumber\\
&&\beta X_1\left(a,\psi,\,\omega t\right)=\beta\sum\limits_{n,m}\gamma_{nm}(a)e^{i(n\omega t+m\psi)}.
\end{eqnarray}
We exclude the secular terms in Eq.(\ref{Bogoliubov4}) and finally deduce:
\begin{widetext}
\begin{eqnarray}\label{Bogoliubov5}
&&\beta f_1=-\frac{\beta}{4\pi^2\frac{p}{q}\omega}\sum\limits_n\exp\left(-in\frac{q}{p}\theta\right)\int\limits_0^{2\pi}\int\limits_0^{2\pi}\exp\left(in\frac{q}{p}\theta\right) Q_0(a,\psi,\omega t)\sin\psi d\psi d(\omega t),\nonumber\\
&&\beta\omega_1=\frac{\beta\Delta}{2\frac{p}{q}a\omega}-\frac{\beta}{4\pi^2\frac{p}{q}\omega}\sum\limits_n\exp\left(-in\frac{q}{p}\theta\right)\int\limits_0^{2\pi}\int\limits_0^{2\pi} \exp\left(in\frac{q}{p}\theta\right) Q_0(a,\psi,\omega t)\cos\psi d\psi d(\omega t).
\end{eqnarray}
\end{widetext}
Taking into account the explicit form of $Q_0(a,\psi,\omega t)$ from Eq.(\ref{Bogoliubov5}) after cumbersome calculations we obtain:
\begin{eqnarray}\label{Bogoliubov6}
&&\beta f_1=-\frac{1}{4\pi^2\omega}\left[2\pi^2\alpha_G a\omega+2\pi^2V_0\cos\theta\right],\nonumber\\
&& \beta\omega_1=\frac{\beta\Delta}{2\omega}+\frac{3a^2\beta}{8\omega}-\frac{V_0}{2a\omega}\sin\theta,
\end{eqnarray}
and 
\begin{eqnarray}\label{Bogoliubov7}
&& \dot{a}=-\frac{\alpha_G}{2}a-\frac{V_0}{2\omega}\cos\theta,\nonumber\\
&& \dot{\theta}=\frac{\beta\Delta}{2\omega}+\frac{3\beta a^2}{8\omega}+\frac{V_0}{2a}\sin\theta.
\end{eqnarray}
Taking into account equilibrium condition $\dot{a}=0$, $\dot{\theta}=0$ we obtain amplitude frequency characteristics of the ferroelectric subsystem:
\begin{eqnarray}\label{Bogoliubov8}
a^2\left[\left(\frac{3\beta}{8}a^2+\frac{\omega_0^2-\omega^2}{2}\right)^2+\frac{\omega^2\alpha^2_G}{4} \right]=\frac{V_0^2}{4}.
\end{eqnarray}
From Eq.(\ref{Bogoliubov8}) we can find the amplitude of oscillation of the ferroelectric subsystem and solve the magnetic subsystem:
\begin{eqnarray}\label{Bogoliubov9}
\ddot{\varphi}+\alpha_G\dot{\varphi}+\left[ 1-ag_{ME}\cos(\omega t+\theta)\right] \varphi=0.
\end{eqnarray}
Analysis of Eq.(\ref{Bogoliubov9}) will be done in the next section.

\section{Magnetoelectric parametric resonance}

We introduce the notation $2\tau=\omega t+\theta$ and rewrite Eq.(\ref{Bogoliubov9}) in the following form:
\begin{eqnarray}\label{Mathieu1}
\ddot{\varphi}+2\gamma_M\dot{\varphi}+\left(\delta^2+\mu^2+\varepsilon\cos2\tau\right) \varphi=0.
\end{eqnarray}
Eq.(\ref{Mathieu1}) is the essence of the Mathieu equation \cite{PhysRevE.71.056211} under the Floquet  ansatz $u=e^{\mu t}\varphi(t)$:
\begin{eqnarray}\label{Mathieu2}
\ddot{u}+(\delta-\varepsilon\cos2\tau)u=0,
\end{eqnarray}
where $\delta=\frac{4}{\omega^2}\left(1-\alpha_G\right)$, $\varepsilon=\frac{4ag_{ME}}{\omega^2}$, $\mu=\frac{2\alpha_G}{\omega^2}$. 
Eq.(\ref{Mathieu2}) has two types of solutions: non-periodic solutions are characterized by parametric magnetoelectric resonance, enhancement of the magnetic osculations through the ferroelectric oscillations. Non-periodic solutions we deduce though the perturbation theory exploiting the 
ansatz $\varphi=\varphi_0+\varepsilon\varphi_1+\varepsilon^2\varphi_2$, $\delta=\delta_0+\varepsilon\delta_1+\varepsilon^2\delta_2$, and $\mu=\varepsilon\mu_1+\varepsilon^2\mu_2$. After cumbersome calculations, one deduces: 
\begin{eqnarray}\label{Mathieu3}
u=Ae^{\left(\frac{\sin2\sigma}{4}\right)t}\left[\sin(t-\sigma)+\frac{1}{16}\varepsilon\sin(3t-\sigma)\right]. 
\end{eqnarray}
Here $\sigma=\arccos\big(\frac{4(1-\alpha_G)-\omega^2}{2ag_{ME}}\big)$. 
The solution of the Mathieu equation is periodic only if the Floquet exponent is imaginary. The periodic solution implies a specific relation between $\delta$ and $\varepsilon$. For each particular periodic Mathieu function $ce_n(\varepsilon, \tau)$ we have a corresponding Mathieu characteristics $A_n\left[ \varepsilon\right] $:
\begin{eqnarray}\label{Mathieu4}
&& \frac{4}{\omega^2}(1-\alpha_G)=A_n\left[ \frac{4a'(\omega)g_{ME}}{\omega^2}\right] ,
\end{eqnarray}
where $a'(\omega)$ is the reals real root of the equation Eq.(\ref{Bogoliubov8})
After solving Eq.(\ref{Bogoliubov8}), Eq.(\ref{Mathieu4}) we can find the frequency of external driving $\omega$ for which dynamics of the magnetic subsystem is periodic. 
For further analysis, we adopt the asymptotic expressions for Mathieu characteristics $A_n(\varepsilon)=n^2+\mathcal{O}(|\varepsilon|^n),\,\varepsilon<1$, $A_n(\varepsilon)=-2\varepsilon+2(2n+1)\sqrt{\varepsilon},\,\varepsilon>1$. The periodic solution reads:
\begin{eqnarray}\label{Mathieu5}
&&X(\omega,\tau)=a'(\omega)\cos(2\tau),\nonumber\\
&& \varphi(t)=\text{ce}_n(\varepsilon, \tau),\nonumber\\
&& \varepsilon=\frac{4a'(\omega)g_{ME}}{\omega^2}.
\end{eqnarray}
We note that the solution is periodic only along the Mathieu characteristics in the parametric space Eq.(\ref{Mathieu4}).

\section{Magneto-Electric coupling and fractals}
We proceed with the analysis of inherently nonlinear limit and consider the effect of a weak dissipation as well. 
We numerically integrate equations of the ME-coupled ferroelectric-nanomagnetic system
\begin{eqnarray}\label{Chaos case}
&&\frac{d^2X}{dt^2}+\alpha_G\frac{dX}{dt}+\gamma X+\beta X^3-g_{ME}\cos\varphi=V_0\sin(\omega t),\nonumber\\
&&\frac{d^2\varphi}{d\tau^2}+\alpha_G\frac{d\varphi}{d\tau}=-\sin\varphi-g_{ME}X\sin\varphi.
\end{eqnarray}
Our interest concerns the formation of magneto-ferroelectric fractal structures in the system and the role of ME coupling in particular. We consider two different types of the confinement potential 
$\gamma<0$ and $\gamma>0$ and plot Poincare sections projecting the four-dimensional phase space of the system on a two-dimensional subspace of the ferroelectric part. 
\begin{figure}[!ht]
\centering
\includegraphics[width=\columnwidth]{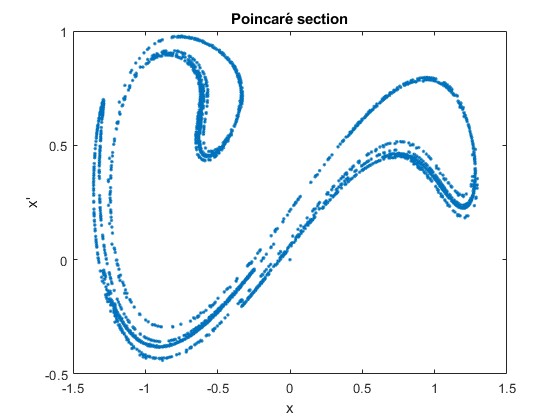}
\caption{Poincare section of the system plotted through projecting four-dimensional phase space of the entire system on a two-dimensional subspace of ferroelectric subsystem. The values of the parameters read: $\gamma = -1, \beta = 1, g_{ME} = 0, \alpha_G = 0.3, V_0 = 0.5, \omega = 1.2.$}
\label{fig:Poincare no ME double-well}
\end{figure}
In Fig.\ref{fig:Poincare no ME double-well} is plotted Poincare section of the ferroelectric subsystem confined in the double-well potential $\gamma<0$ in the absence of the ME coupling. 
As we see phase trajectory is winded on two equilibrium points $X=-1$, $X=1$ characterizing the double well-potential. 

\begin{figure}[!ht]
\centering
\includegraphics[width=\columnwidth]{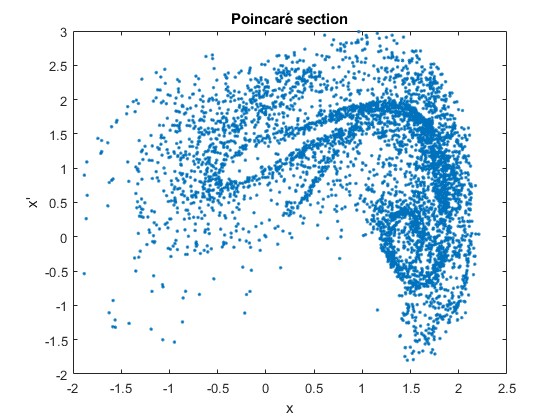}
\caption{Poincare section of the system plotted through projecting four-dimensional phase space of the entire system on the two-dimensional subspace of ferroelectric subsystem. The values of the parameters read: $\gamma = -1, \beta = 1, g_{ME} = 1, \alpha_G = 0.3, V_0 = 0.5, \omega = 1.2.$}
\label{fig:Poincare with ME double-well}
\end{figure}

We switch on to the discussion of ME coupling and plot the phase portrait in Fig.\ref{fig:Poincare with ME double-well}. As we see chaos sets in the system and in the case of the double-well potential 
the system is too sensitive with respect to the ME coupling term $g_{ME}$. This numerical result is in line with the theoretical analysis done above. 

\begin{figure}[!ht]
\centering
\includegraphics[width=\columnwidth]{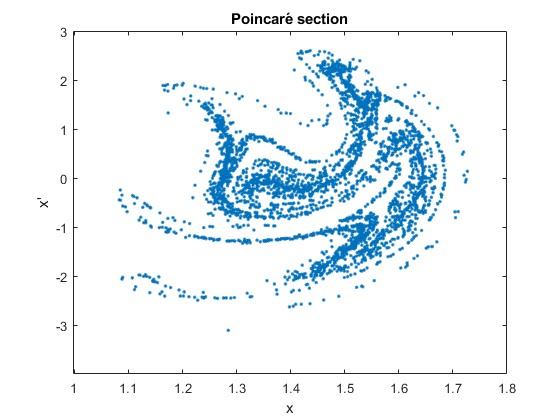}
\caption{Poincare section of the system plotted through projecting four-dimensional phase space of the entire system on the two-dimensional subspace of ferroelectric subsystem. The values of the parameters read: $\gamma = 1, \beta = 5, g_{ME} = 0, \alpha_G = 0.02, V_0 = 8, \omega = 0.5.$}
\label{fig:Poincare single well no ME}
\end{figure}

\begin{figure}[!ht]
\centering
\includegraphics[width=\columnwidth]{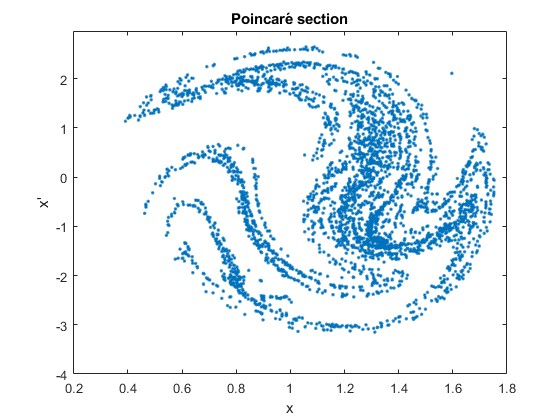}
\caption{Poincare section of the system plotted through projecting four-dimensional phase space of the entire system on the two-dimensional subspace of ferroelectric subsystem. 
The values of the parameters read: $\gamma = 1, \beta = 5, g_{ME} = 0.5, \alpha_G = 0.02, V_0 = 8, \omega = 0.5.$}
\label{fig:Poincare single well with ME}
\end{figure}

We proceed with the single well confinement potential. Results are plotted in Fig.\ref{fig:Poincare single well no ME} and  Fig.\ref{fig:Poincare single well with ME}. As we see, in the case of single well potential $\gamma>0$ the ferroelectric system is less sensitive to the ME term.  To quantify the chaos we study Lyapunov functions and the Fractal dimension in the system. 
We define the vector of the state $\textbf{x}(t)=[X(t),\dot{X}(t),\varphi(t),\dot{\varphi} (t)]$ and the initial small distance between two phase trajectories $\delta_0=|\textbf{x}(0)-\textbf{x}'(0)|$. Then the maximal Lyapunov exponent is given by
\begin{eqnarray}
\lambda(g_{ME})=\lim\limits_{N \to \infty}\frac{1}{(N+1)\Delta t}\sum\limits_{n=0}^{N}\log\bigg\Vert\frac{\delta x_n}{|\delta x_{n-1}|}\bigg\Vert.
\end{eqnarray}
Here $\Delta t$ is the time step and the mean value of the exponent is calculated over the set of initial several hundred trajectories. For exploring the Fractal dimension we utilize the Grassberger–Procaccia proposal \cite{PhysRevLett.50.346}
\begin{widetext}
\begin{eqnarray}
C(\varepsilon)=\lim\limits_{N\to\infty}\frac{1}{N(N-1)}\sum\limits_{i,j=1}^N\theta(\varepsilon-|\textbf{x}(t_i)-\textbf{x}(t_j)|).
\end{eqnarray}
\end{widetext}
Here $\theta$ is the Heaviside step function. We plot $\log C(\varepsilon)$ as a function of $\log(\varepsilon)$. The angular coefficient of the linear regression of the graph yields $D$. 

\begin{table}[ht]
    \centering
    \begin{tabular}{|c|c|c|c|c|c|c|c|c|}
    \hline
    $g_{ME}$ & 0.00 & 0.07 & 0.30 & 0.43 & 0.61 & 0.76 & 0.90 & 1.00 \\ \hline
    $\lambda$ & 0.014 & 0.022 & 0.020 & 0.000 & 0.000 & 0.001 & 0.035 & 0.030 \\ \hline
    $C$ & 1.48 & 1.34 & 1.46 & 0.06 & 0.02 & 0.09 & 1.83 & 2.02 \\ \hline
    \end{tabular}
    \caption{Lyapunov exponent ($\lambda$) and fractal dimension ($C$) for various values of magnetoelectric coupling ($g_{ME}$) for parameters: $\gamma = -1, \beta = 1,  \alpha_G = 0.3, V_0 = 0.5, \omega = 1.2.$}
    \label{tab:case_1}
\end{table}

\begin{table}[ht]
    \centering
    \begin{tabular}{|c|c|c|c|c|}
    \hline
    $g_{ME}$ & 0.00 & 0.20 & 0.56 & 1.00 \\ \hline
    $\lambda$ & 0.019 & 0.022 & 0.018 & 0.038  \\ \hline
    $C$ & 2.25 & 2.11 & 2.28 & 1.37  \\ \hline
    \end{tabular}
    \caption{Lyapunov exponent ($\lambda$) and fractal dimension ($C$) for various values of magnetoelectric coupling ($g_{ME}$) for parameters: $\gamma = 1, \beta = 5,  \alpha_G = 0.02, V_0 = 8, \omega = 0.5.$}
    \label{tab:case_2}
\end{table}

\begin{table}[ht]
    \centering
    \begin{tabular}{|c|c|c|c|c|}
    \hline
    $g_{ME}$ & 0.00 & 0.50 & 0.75 & 1.00 \\ \hline
    $\lambda$ & -0.002 & -0.006 & 0.025 & 0.031  \\ \hline
    $C$ & 0.06 & 0.05 & 1.7 & 1.95  \\ \hline
    \end{tabular}
    \caption{Lyapunov exponent ($\lambda$) and fractal dimension ($C$) for various values of magnetoelectric coupling ($g_{ME}$) for parameters: $\gamma = -1, \beta = 0.5,  \alpha_G = 0.3, V_0 = 0.2, \omega = 1.2.$}
    \label{tab:case_3}
\end{table}

\begin{figure*}[!ht]
\centering
\includegraphics[width=\textwidth]{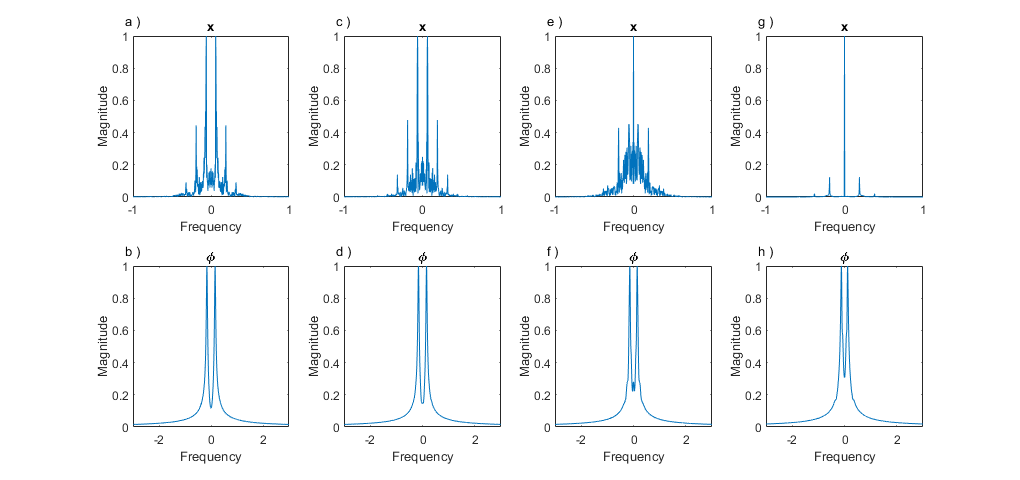}
\caption{Plots of fast Fourier transform for $x$ and $\phi$ for different values of $g_{ME}$. The values of the parameters read: $\gamma = -1, \beta = 1,  \alpha_G = 0.3, V_0 = 0.5, \omega = 1.2.$. Plots a) for $x$ and b) for $\phi$ both for $g_{ME} = 0.0$, c) for $x$ and d) for $\phi$ both for $g_{ME} = 0.07$, e) for $x$ and f) for $\phi$ both for $g_{ME} = 0.30$, g) for $x$ and h) for $\phi$ both for $g_{ME} = 0.43$,}
\label{fig:ffts plot for table 1 part 1}
\end{figure*}

\begin{figure*}[!ht]
\centering
\includegraphics[width=\textwidth]{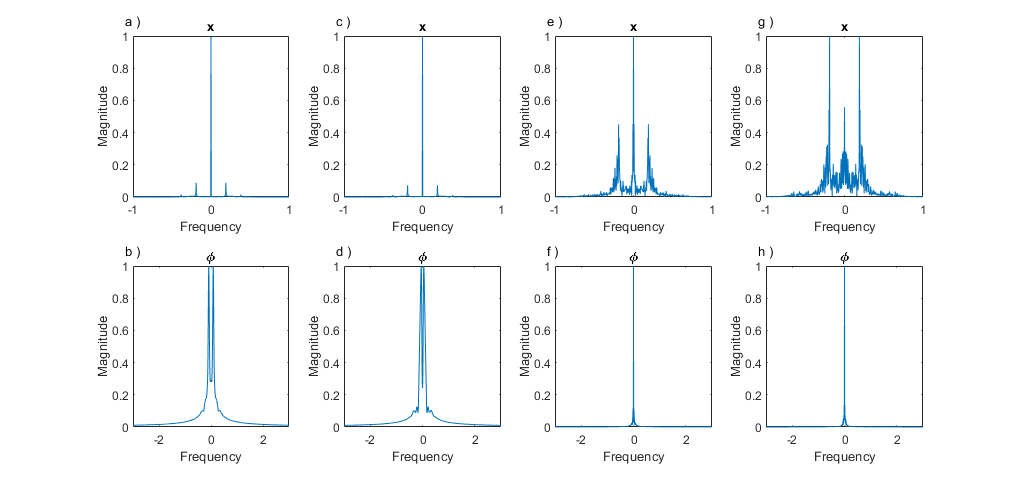}
\caption{Plots of fast Fourier transform for $x$ and $\phi$ for different values of $g_{ME}$. The values of the parameters read: $\gamma = -1, \beta = 1,  \alpha_G = 0.3, V_0 = 0.5, \omega = 1.2.$. Plots a) for $x$ and b) for $\phi$ both for $g_{ME} = 0.61$, c) for $x$ and d) for $\phi$ both for $g_{ME} = 0.76$, e) for $x$ and f) for $\phi$ both for $g_{ME} = 0.90$, g) for $x$ and h) for $\phi$ both for $g_{ME} = 1.0$,}
\label{fig:ffts plot for table 1 part 2}
\end{figure*}

\begin{figure*}[!ht]
\centering
\includegraphics[width=\textwidth]{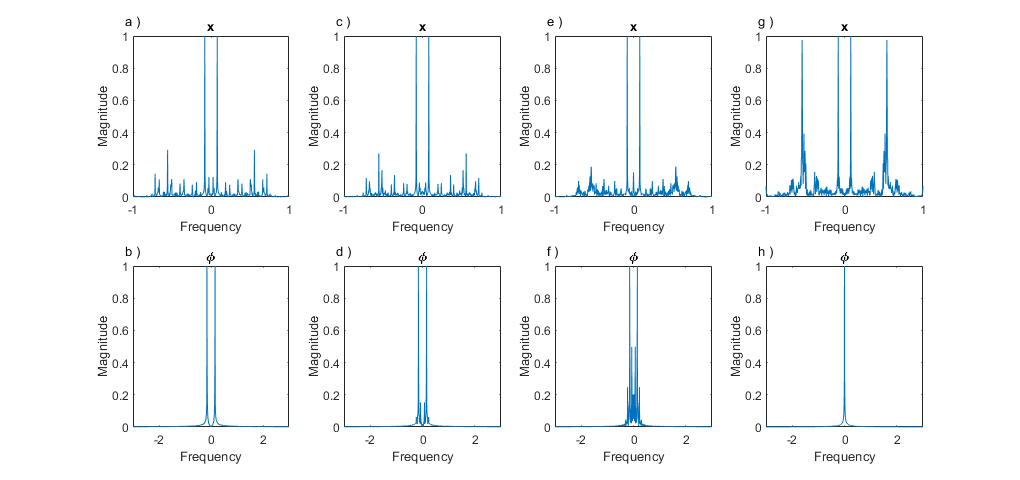}
\caption{Plots of fast Fourier transform for $x$ and $\phi$ for different values of $g_{ME}$. The values of the parameters read: $\gamma = 1, \beta = 5,  \alpha_G = 0.02, V_0 = 8, \omega = 0.5.$. Plots a) for $x$ and b) for $\phi$ both for $g_{ME} = 0.0$, c) for $x$ and d) for $\phi$ both for $g_{ME} = 0.20$, e) for $x$ and f) for $\phi$ both for $g_{ME} = 0.56$, g) for $x$ and h) for $\phi$ both for $g_{ME} = 1.0$,}
\label{fig:ffts plot for table 2}
\end{figure*}

\begin{figure*}[!ht]
\centering
\includegraphics[width=\textwidth]{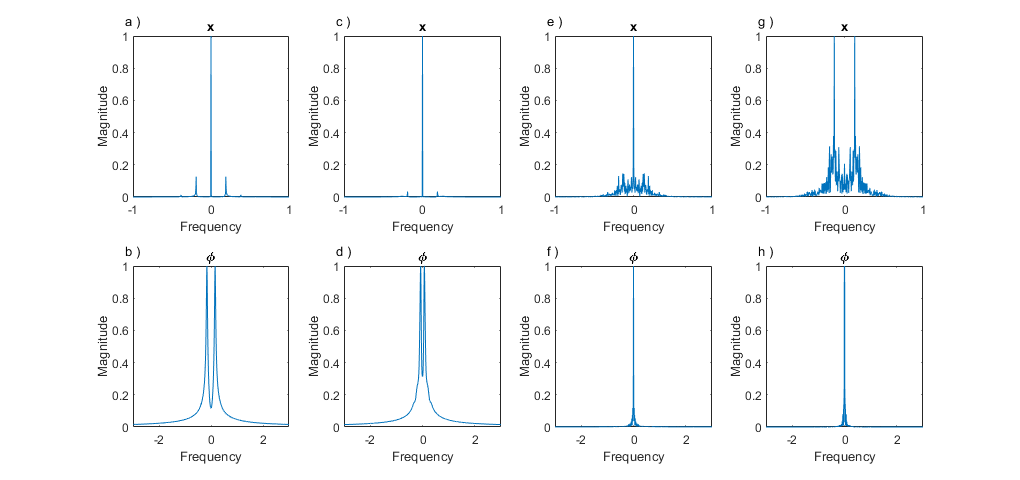}
\caption{Plots of fast Fourier transform for $x$ and $\phi$ for different values of $g_{ME}$. The values of the parameters read: $\gamma = -1, \beta = 0.5,  \alpha_G = 0.3, V_0 = 0.2, \omega = 1.2.$. Plots a) for $x$ and b) for $\phi$ both for $g_{ME} = 0.0$, c) for $x$ and d) for $\phi$ both for $g_{ME} = 0.50$, e) for $x$ and f) for $\phi$ both for $g_{ME} = 0.75$, g) for $x$ and h) for $\phi$ both for $g_{ME} = 1.0$,}
\label{fig:ffts plot for table 3}
\end{figure*}

We calculate Lyapunov exponents and Fractal dimensions for different values of ME coupling constant $g_{ME}$. We consider both cases of the positive and negative $\gamma$ corresponding to the different forms of ferroelectric confinement potentials. The results of the calculations are shown in Tables I-III. We clearly see the nontrivial role of the ME coupling. When starting from the $g_{ME}=0$ and the chaotic initial state $\lambda>$, (Table I), we gradually increase the value of ME coupling. The stronger chaos we observe for large $g_{ME}$. However, there is a region of the regular dynamics for particular values of ME coupling $0<g_{ME}<1$. In the case of the zero or negative Lyapunov exponents, the fractal dimension approaches the integer number (Table I, Table III). For further insights, we study the Furrier spectrum of individual subsystems
Fig.\ref{fig:ffts plot for table 1 part 1}, Fig.\ref{fig:ffts plot for table 1 part 2}, Fig.\ref{fig:ffts plot for table 2} and Fig.\ref{fig:ffts plot for table 3}. The broadening of the spectrum is a hallmark of the chaos in the particular subsystem. There are cases when both subsystems display chaos (finite width of the spectrums) and cases when one subsystem is regular. An interesting phenomenon occurs when steering the value of ME coupling. In Fig.\ref{fig:ffts plot for table 1 part 1}, the ferroelectric system is chaotic for a small ME coupling. For another set of parameters shown in Fig.\ref{fig:ffts plot for table 1 part 2}, chaos in the ferroelectric system is stronger when ME coupling is large. The magnetic subsystem shows the opposite behavior. Thus we can conclude that chaotic behavior is transferred between the subsystems. 

\section{Conclusions}

The magnetoelectric coupling effect has attracted vast interest during the last decades. The mechanism of the magnetoelectric coupling at the ferroelectric-magnetic interface is based on screening effects, and it influences both the ferroelectric and magnetic properties of the system. 
Different aspects of the coupled ferroelectric and magnetic systems were studied intensively. However, nonlinear dynamic aspects of the coupled ferroelectric crystal-magnetic nanoparticle system were not addressed in a thoroughly general and rigorous mathematical form. In the present work, we tried to fill this gap. We showed that by applying two time-dependent electric fields with different frequencies, we could dynamically design the confinement potential of the ferroelectric subsystem and change its shape from quintic to double-well potential. This fact allowed us to realize two different types of dynamics. Concerning the ME coupling, we showed that the system is more sensitive to it in the case of double-well potential. We started our study with the Hamiltonian approach and addressed two moderate and strong nonlinearity cases. 
We dedicated our discussion to low and zero temperatures cases and eliminated the factors of thermal environment and noise. For certain materials, e.g., YIG, the Gilbert damping factor is small, and therefore, the Hamiltonian method is a valid physical approach, at least for a short period of dynamics. We implemented the Kolmogorov Arnold Moser theorem, utilized canonical action-angle variables, and studied the overlapping of nonlinear resonances. We found two invariant tors of the hybrid system in the system's phase space. The ME coupling term in our discussion played a role of the small perturbation destroying invariant tours  (in accordance with the KAM theorem). Analyses of Melnikov's function showed that in the limit of moderate nonlinearity, the dynamic of the magnetic subsystem is chaotic, while the dynamic of the ferroelectric subsystem is regular. In the case of strong nonlinearity, analysis of the phase space region near the separatrix showed the formation of a homoclinic structure, and the dynamic of both ferroelectric and magnetic subsystems is chaotic even for an arbitrary small ME coupling term. 
We analyzed the character of bifurcations in the system and found that those are of Hopf's type. They occur when tuning the amplitude of the ME coupling term. We also studied the problem of weak nonlinearity. We assumed that the deviation of the system from equilibrium is relatively tiny and implemented Van der Pol's method in the non-resonant case. In contrast, in the resonant case, we implemented Bogoliubov's method. We linearized the system near a particular nonlinear resonance and explored the overlapping of the nonlinear resonances. We found that the parametric resonance problem describes the magnetic subsystem's dynamic. Besides, we explored the problem of parametric resonance and the possibility of enhancing magnetic oscillations through the ferroelectric subsystem. In analyses of the Mathieu equation, we 
discovered particular values of parameters when the dynamic of the magnetic system becomes periodic in time. All these conditions are experimentally feasible. We also analyzed the problem of strong nonlinearity and dissipation processes in the system. We studied Lyapunov's function and fractal dimension and found a strong dependence of both quantities on the ME coupling term. 

\section*{acknowledgement}
We thank Wiktoria Wojnarowska for fruitful discussions. 
 

\bibliography{TEXT_KK_VD.LC}

\begin{thebibliography}{37}%
\makeatletter
\providecommand \@ifxundefined [1]{%
 \@ifx{#1\undefined}
}%
\providecommand \@ifnum [1]{%
 \ifnum #1\expandafter \@firstoftwo
 \else \expandafter \@secondoftwo
 \fi
}%
\providecommand \@ifx [1]{%
 \ifx #1\expandafter \@firstoftwo
 \else \expandafter \@secondoftwo
 \fi
}%
\providecommand \natexlab [1]{#1}%
\providecommand \enquote  [1]{``#1''}%
\providecommand \bibnamefont  [1]{#1}%
\providecommand \bibfnamefont [1]{#1}%
\providecommand \citenamefont [1]{#1}%
\providecommand \href@noop [0]{\@secondoftwo}%
\providecommand \href [0]{\begingroup \@sanitize@url \@href}%
\providecommand \@href[1]{\@@startlink{#1}\@@href}%
\providecommand \@@href[1]{\endgroup#1\@@endlink}%
\providecommand \@sanitize@url [0]{\catcode `\\12\catcode `\$12\catcode
  `\&12\catcode `\#12\catcode `\^12\catcode `\_12\catcode `\%12\relax}%
\providecommand \@@startlink[1]{}%
\providecommand \@@endlink[0]{}%
\providecommand \url  [0]{\begingroup\@sanitize@url \@url }%
\providecommand \@url [1]{\endgroup\@href {#1}{\urlprefix }}%
\providecommand \urlprefix  [0]{URL }%
\providecommand \Eprint [0]{\href }%
\providecommand \doibase [0]{https://doi.org/}%
\providecommand \selectlanguage [0]{\@gobble}%
\providecommand \bibinfo  [0]{\@secondoftwo}%
\providecommand \bibfield  [0]{\@secondoftwo}%
\providecommand \translation [1]{[#1]}%
\providecommand \BibitemOpen [0]{}%
\providecommand \bibitemStop [0]{}%
\providecommand \bibitemNoStop [0]{.\EOS\space}%
\providecommand \EOS [0]{\spacefactor3000\relax}%
\providecommand \BibitemShut  [1]{\csname bibitem#1\endcsname}%
\let\auto@bib@innerbib\@empty
\bibitem [{\citenamefont {Farzin}\ \emph {et~al.}(2020)\citenamefont {Farzin},
  \citenamefont {Etesami}, \citenamefont {Quint}, \citenamefont {Memic},\ and\
  \citenamefont {Tamayol}}]{farzin2020magnetic}%
  \BibitemOpen
  \bibfield  {author} {\bibinfo {author} {\bibfnamefont {A.}~\bibnamefont
  {Farzin}}, \bibinfo {author} {\bibfnamefont {S.~A.}\ \bibnamefont {Etesami}},
  \bibinfo {author} {\bibfnamefont {J.}~\bibnamefont {Quint}}, \bibinfo
  {author} {\bibfnamefont {A.}~\bibnamefont {Memic}},\ and\ \bibinfo {author}
  {\bibfnamefont {A.}~\bibnamefont {Tamayol}},\ }\bibfield  {title} {\bibinfo
  {title} {Magnetic nanoparticles in cancer therapy and diagnosis},\
  }\href@noop {} {\bibfield  {journal} {\bibinfo  {journal} {Advanced
  healthcare materials}\ }\textbf {\bibinfo {volume} {9}},\ \bibinfo {pages}
  {1901058} (\bibinfo {year} {2020})}\BibitemShut {NoStop}%
\bibitem [{\citenamefont {Gloag}\ \emph {et~al.}(2019)\citenamefont {Gloag},
  \citenamefont {Mehdipour}, \citenamefont {Chen}, \citenamefont {Tilley},\
  and\ \citenamefont {Gooding}}]{gloag2019advances}%
  \BibitemOpen
  \bibfield  {author} {\bibinfo {author} {\bibfnamefont {L.}~\bibnamefont
  {Gloag}}, \bibinfo {author} {\bibfnamefont {M.}~\bibnamefont {Mehdipour}},
  \bibinfo {author} {\bibfnamefont {D.}~\bibnamefont {Chen}}, \bibinfo {author}
  {\bibfnamefont {R.~D.}\ \bibnamefont {Tilley}},\ and\ \bibinfo {author}
  {\bibfnamefont {J.~J.}\ \bibnamefont {Gooding}},\ }\bibfield  {title}
  {\bibinfo {title} {Advances in the application of magnetic nanoparticles for
  sensing},\ }\href@noop {} {\bibfield  {journal} {\bibinfo  {journal}
  {Advanced Materials}\ }\textbf {\bibinfo {volume} {31}},\ \bibinfo {pages}
  {1904385} (\bibinfo {year} {2019})}\BibitemShut {NoStop}%
\bibitem [{\citenamefont {Shasha}\ and\ \citenamefont
  {Krishnan}(2021)}]{shasha2021nonequilibrium}%
  \BibitemOpen
  \bibfield  {author} {\bibinfo {author} {\bibfnamefont {C.}~\bibnamefont
  {Shasha}}\ and\ \bibinfo {author} {\bibfnamefont {K.~M.}\ \bibnamefont
  {Krishnan}},\ }\bibfield  {title} {\bibinfo {title} {Nonequilibrium dynamics
  of magnetic nanoparticles with applications in biomedicine},\ }\href@noop {}
  {\bibfield  {journal} {\bibinfo  {journal} {Advanced Materials}\ }\textbf
  {\bibinfo {volume} {33}},\ \bibinfo {pages} {1904131} (\bibinfo {year}
  {2021})}\BibitemShut {NoStop}%
\bibitem [{\citenamefont {Liu}\ \emph {et~al.}(2020)\citenamefont {Liu},
  \citenamefont {Yu}, \citenamefont {Wang}, \citenamefont {Shen},\ and\
  \citenamefont {Cong}}]{liu2020preparation}%
  \BibitemOpen
  \bibfield  {author} {\bibinfo {author} {\bibfnamefont {S.}~\bibnamefont
  {Liu}}, \bibinfo {author} {\bibfnamefont {B.}~\bibnamefont {Yu}}, \bibinfo
  {author} {\bibfnamefont {S.}~\bibnamefont {Wang}}, \bibinfo {author}
  {\bibfnamefont {Y.}~\bibnamefont {Shen}},\ and\ \bibinfo {author}
  {\bibfnamefont {H.}~\bibnamefont {Cong}},\ }\bibfield  {title} {\bibinfo
  {title} {Preparation, surface functionalization and application of fe3o4
  magnetic nanoparticles},\ }\href@noop {} {\bibfield  {journal} {\bibinfo
  {journal} {Advances in colloid and Interface Science}\ }\textbf {\bibinfo
  {volume} {281}},\ \bibinfo {pages} {102165} (\bibinfo {year}
  {2020})}\BibitemShut {NoStop}%
\bibitem [{\citenamefont {Wu}\ \emph {et~al.}(2019)\citenamefont {Wu},
  \citenamefont {Su}, \citenamefont {Liu}, \citenamefont {Saha},\ and\
  \citenamefont {Wang}}]{wu2019magnetic}%
  \BibitemOpen
  \bibfield  {author} {\bibinfo {author} {\bibfnamefont {K.}~\bibnamefont
  {Wu}}, \bibinfo {author} {\bibfnamefont {D.}~\bibnamefont {Su}}, \bibinfo
  {author} {\bibfnamefont {J.}~\bibnamefont {Liu}}, \bibinfo {author}
  {\bibfnamefont {R.}~\bibnamefont {Saha}},\ and\ \bibinfo {author}
  {\bibfnamefont {J.-P.}\ \bibnamefont {Wang}},\ }\bibfield  {title} {\bibinfo
  {title} {Magnetic nanoparticles in nanomedicine: a review of recent
  advances},\ }\href@noop {} {\bibfield  {journal} {\bibinfo  {journal}
  {Nanotechnology}\ }\textbf {\bibinfo {volume} {30}},\ \bibinfo {pages}
  {502003} (\bibinfo {year} {2019})}\BibitemShut {NoStop}%
\bibitem [{\citenamefont {Hubert}\ \emph {et~al.}(2021)\citenamefont {Hubert},
  \citenamefont {Trosman}, \citenamefont {Collard}, \citenamefont {Sukhov},
  \citenamefont {Harting}, \citenamefont {Vandewalle},\ and\ \citenamefont
  {Smith}}]{PhysRevLett.126.224501}%
  \BibitemOpen
  \bibfield  {author} {\bibinfo {author} {\bibfnamefont {M.}~\bibnamefont
  {Hubert}}, \bibinfo {author} {\bibfnamefont {O.}~\bibnamefont {Trosman}},
  \bibinfo {author} {\bibfnamefont {Y.}~\bibnamefont {Collard}}, \bibinfo
  {author} {\bibfnamefont {A.}~\bibnamefont {Sukhov}}, \bibinfo {author}
  {\bibfnamefont {J.}~\bibnamefont {Harting}}, \bibinfo {author} {\bibfnamefont
  {N.}~\bibnamefont {Vandewalle}},\ and\ \bibinfo {author} {\bibfnamefont
  {A.-S.}\ \bibnamefont {Smith}},\ }\bibfield  {title} {\bibinfo {title}
  {Scallop theorem and swimming at the mesoscale},\ }\href
  {https://doi.org/10.1103/PhysRevLett.126.224501} {\bibfield  {journal}
  {\bibinfo  {journal} {Phys. Rev. Lett.}\ }\textbf {\bibinfo {volume} {126}},\
  \bibinfo {pages} {224501} (\bibinfo {year} {2021})}\BibitemShut {NoStop}%
\bibitem [{\citenamefont {Toklikishvili}\ \emph {et~al.}(2023)\citenamefont
  {Toklikishvili}, \citenamefont {Chotorlishvili}, \citenamefont {Khomeriki},
  \citenamefont {Jandieri},\ and\ \citenamefont
  {Berakdar}}]{PhysRevB.107.115126}%
  \BibitemOpen
  \bibfield  {author} {\bibinfo {author} {\bibfnamefont {Z.}~\bibnamefont
  {Toklikishvili}}, \bibinfo {author} {\bibfnamefont {L.}~\bibnamefont
  {Chotorlishvili}}, \bibinfo {author} {\bibfnamefont {R.}~\bibnamefont
  {Khomeriki}}, \bibinfo {author} {\bibfnamefont {V.}~\bibnamefont
  {Jandieri}},\ and\ \bibinfo {author} {\bibfnamefont {J.}~\bibnamefont
  {Berakdar}},\ }\bibfield  {title} {\bibinfo {title} {Electrically controlled
  entanglement of cavity photons with electromagnons},\ }\href
  {https://doi.org/10.1103/PhysRevB.107.115126} {\bibfield  {journal} {\bibinfo
   {journal} {Phys. Rev. B}\ }\textbf {\bibinfo {volume} {107}},\ \bibinfo
  {pages} {115126} (\bibinfo {year} {2023})}\BibitemShut {NoStop}%
\bibitem [{\citenamefont {Khomeriki}\ \emph {et~al.}(2015)\citenamefont
  {Khomeriki}, \citenamefont {Chotorlishvili}, \citenamefont {Malomed},\ and\
  \citenamefont {Berakdar}}]{PhysRevB.91.041408}%
  \BibitemOpen
  \bibfield  {author} {\bibinfo {author} {\bibfnamefont {R.}~\bibnamefont
  {Khomeriki}}, \bibinfo {author} {\bibfnamefont {L.}~\bibnamefont
  {Chotorlishvili}}, \bibinfo {author} {\bibfnamefont {B.~A.}\ \bibnamefont
  {Malomed}},\ and\ \bibinfo {author} {\bibfnamefont {J.}~\bibnamefont
  {Berakdar}},\ }\bibfield  {title} {\bibinfo {title} {Creation and
  amplification of electromagnon solitons by electric field in nanostructured
  multiferroics},\ }\href {https://doi.org/10.1103/PhysRevB.91.041408}
  {\bibfield  {journal} {\bibinfo  {journal} {Phys. Rev. B}\ }\textbf {\bibinfo
  {volume} {91}},\ \bibinfo {pages} {041408} (\bibinfo {year}
  {2015})}\BibitemShut {NoStop}%
\bibitem [{\citenamefont {Khomeriki}\ \emph {et~al.}(2016)\citenamefont
  {Khomeriki}, \citenamefont {Chotorlishvili}, \citenamefont {Tralle},\ and\
  \citenamefont {Berakdar}}]{khomeriki2016positive}%
  \BibitemOpen
  \bibfield  {author} {\bibinfo {author} {\bibfnamefont {R.}~\bibnamefont
  {Khomeriki}}, \bibinfo {author} {\bibfnamefont {L.}~\bibnamefont
  {Chotorlishvili}}, \bibinfo {author} {\bibfnamefont {I.}~\bibnamefont
  {Tralle}},\ and\ \bibinfo {author} {\bibfnamefont {J.}~\bibnamefont
  {Berakdar}},\ }\bibfield  {title} {\bibinfo {title} {Positive--negative
  birefringence in multiferroic layered metasurfaces},\ }\href@noop {}
  {\bibfield  {journal} {\bibinfo  {journal} {Nano Letters}\ }\textbf {\bibinfo
  {volume} {16}},\ \bibinfo {pages} {7290} (\bibinfo {year}
  {2016})}\BibitemShut {NoStop}%
\bibitem [{\citenamefont {Choudhury}\ \emph {et~al.}(2004)\citenamefont
  {Choudhury}, \citenamefont {Chitra},\ and\ \citenamefont
  {Ramanadham}}]{choudhury2004ferroelectric}%
  \BibitemOpen
  \bibfield  {author} {\bibinfo {author} {\bibfnamefont {R.~R.}\ \bibnamefont
  {Choudhury}}, \bibinfo {author} {\bibfnamefont {R.}~\bibnamefont {Chitra}},\
  and\ \bibinfo {author} {\bibfnamefont {M.}~\bibnamefont {Ramanadham}},\
  }\bibfield  {title} {\bibinfo {title} {Ferroelectric phase transition in
  triglycine selenate: an interpretation based on its structure and its
  comparison with triglycine sulphate},\ }\href@noop {} {\bibfield  {journal}
  {\bibinfo  {journal} {Phase Transitions}\ }\textbf {\bibinfo {volume} {77}},\
  \bibinfo {pages} {385} (\bibinfo {year} {2004})}\BibitemShut {NoStop}%
\bibitem [{\citenamefont {Choudhury}\ and\ \citenamefont
  {Chitra}(2009)}]{choudhury2009structural}%
  \BibitemOpen
  \bibfield  {author} {\bibinfo {author} {\bibfnamefont {R.~R.}\ \bibnamefont
  {Choudhury}}\ and\ \bibinfo {author} {\bibfnamefont {R.}~\bibnamefont
  {Chitra}},\ }\bibfield  {title} {\bibinfo {title} {Structural origin for the
  change of the order of ferroelectric phase transition in triglycine
  sulfate/selenate systems},\ }\href@noop {} {\bibfield  {journal} {\bibinfo
  {journal} {Journal of Physics: Condensed Matter}\ }\textbf {\bibinfo {volume}
  {21}},\ \bibinfo {pages} {335901} (\bibinfo {year} {2009})}\BibitemShut
  {NoStop}%
\bibitem [{\citenamefont {Choudhury}\ \emph {et~al.}(2003)\citenamefont
  {Choudhury}, \citenamefont {Chitra},\ and\ \citenamefont
  {Ramanadham}}]{choudhury2003role}%
  \BibitemOpen
  \bibfield  {author} {\bibinfo {author} {\bibfnamefont {R.~R.}\ \bibnamefont
  {Choudhury}}, \bibinfo {author} {\bibfnamefont {R.}~\bibnamefont {Chitra}},\
  and\ \bibinfo {author} {\bibfnamefont {M.}~\bibnamefont {Ramanadham}},\
  }\bibfield  {title} {\bibinfo {title} {The role of the double-well potential
  seen by the amino group in the ferroelectric phase transition in triglycine
  sulfate},\ }\href@noop {} {\bibfield  {journal} {\bibinfo  {journal} {Journal
  of Physics: Condensed Matter}\ }\textbf {\bibinfo {volume} {15}},\ \bibinfo
  {pages} {4641} (\bibinfo {year} {2003})}\BibitemShut {NoStop}%
\bibitem [{\citenamefont {Trybus}(2020)}]{trybus2020phase}%
  \BibitemOpen
  \bibfield  {author} {\bibinfo {author} {\bibfnamefont {M.}~\bibnamefont
  {Trybus}},\ }\bibfield  {title} {\bibinfo {title} {Phase transition in
  triglycine sulphate investigated using two-phase bridge measurements},\
  }\href@noop {} {\bibfield  {journal} {\bibinfo  {journal} {Infrared Physics
  \& Technology}\ }\textbf {\bibinfo {volume} {109}},\ \bibinfo {pages}
  {103409} (\bibinfo {year} {2020})}\BibitemShut {NoStop}%
\bibitem [{\citenamefont {Trybus}\ \emph {et~al.}(2016)\citenamefont {Trybus},
  \citenamefont {Paszkiewicz},\ and\ \citenamefont
  {Wos}}]{trybus2016observation}%
  \BibitemOpen
  \bibfield  {author} {\bibinfo {author} {\bibfnamefont {M.}~\bibnamefont
  {Trybus}}, \bibinfo {author} {\bibfnamefont {T.}~\bibnamefont
  {Paszkiewicz}},\ and\ \bibinfo {author} {\bibfnamefont {B.}~\bibnamefont
  {Wos}},\ }\bibfield  {title} {\bibinfo {title} {Observation of dynamics of
  hydrogen bonds in tgs crystals by means of measurements of pyroelectric
  currents induced by changes of temperature},\ }\href@noop {} {\bibfield
  {journal} {\bibinfo  {journal} {Infrared Physics \& Technology}\ }\textbf
  {\bibinfo {volume} {79}},\ \bibinfo {pages} {128} (\bibinfo {year}
  {2016})}\BibitemShut {NoStop}%
\bibitem [{\citenamefont {Trybus}\ and\ \citenamefont
  {Wo{\'s}}(2015)}]{trybus2015dynamic}%
  \BibitemOpen
  \bibfield  {author} {\bibinfo {author} {\bibfnamefont {M.}~\bibnamefont
  {Trybus}}\ and\ \bibinfo {author} {\bibfnamefont {B.}~\bibnamefont
  {Wo{\'s}}},\ }\bibfield  {title} {\bibinfo {title} {Dynamic response of tgs
  ferroelectric samples in paraelectric phase},\ }\href@noop {} {\bibfield
  {journal} {\bibinfo  {journal} {Infrared Physics \& Technology}\ }\textbf
  {\bibinfo {volume} {71}},\ \bibinfo {pages} {526} (\bibinfo {year}
  {2015})}\BibitemShut {NoStop}%
\bibitem [{\citenamefont {Rikken}\ and\ \citenamefont
  {Avarvari}(2022{\natexlab{a}})}]{PhysRevB.106.224307}%
  \BibitemOpen
  \bibfield  {author} {\bibinfo {author} {\bibfnamefont {G.~L. J.~A.}\
  \bibnamefont {Rikken}}\ and\ \bibinfo {author} {\bibfnamefont
  {N.}~\bibnamefont {Avarvari}},\ }\bibfield  {title} {\bibinfo {title}
  {Dielectric magnetochiral anisotropy in triglycine sulfate},\ }\href
  {https://doi.org/10.1103/PhysRevB.106.224307} {\bibfield  {journal} {\bibinfo
   {journal} {Phys. Rev. B}\ }\textbf {\bibinfo {volume} {106}},\ \bibinfo
  {pages} {224307} (\bibinfo {year} {2022}{\natexlab{a}})}\BibitemShut
  {NoStop}%
\bibitem [{\citenamefont {Rikken}\ and\ \citenamefont
  {Avarvari}(2022{\natexlab{b}})}]{rikken2022dielectric}%
  \BibitemOpen
  \bibfield  {author} {\bibinfo {author} {\bibfnamefont {G.~L.}\ \bibnamefont
  {Rikken}}\ and\ \bibinfo {author} {\bibfnamefont {N.}~\bibnamefont
  {Avarvari}},\ }\bibfield  {title} {\bibinfo {title} {Dielectric magnetochiral
  anisotropy},\ }\href@noop {} {\bibfield  {journal} {\bibinfo  {journal}
  {Nature Communications}\ }\textbf {\bibinfo {volume} {13}},\ \bibinfo {pages}
  {3564} (\bibinfo {year} {2022}{\natexlab{b}})}\BibitemShut {NoStop}%
\bibitem [{\citenamefont {Chotorlishvili}\ \emph {et~al.}(2013)\citenamefont
  {Chotorlishvili}, \citenamefont {Khomeriki}, \citenamefont {Sukhov},
  \citenamefont {Ruffo},\ and\ \citenamefont
  {Berakdar}}]{PhysRevLett.111.117202}%
  \BibitemOpen
  \bibfield  {author} {\bibinfo {author} {\bibfnamefont {L.}~\bibnamefont
  {Chotorlishvili}}, \bibinfo {author} {\bibfnamefont {R.}~\bibnamefont
  {Khomeriki}}, \bibinfo {author} {\bibfnamefont {A.}~\bibnamefont {Sukhov}},
  \bibinfo {author} {\bibfnamefont {S.}~\bibnamefont {Ruffo}},\ and\ \bibinfo
  {author} {\bibfnamefont {J.}~\bibnamefont {Berakdar}},\ }\bibfield  {title}
  {\bibinfo {title} {Dynamics of localized modes in a composite multiferroic
  chain},\ }\href {https://doi.org/10.1103/PhysRevLett.111.117202} {\bibfield
  {journal} {\bibinfo  {journal} {Phys. Rev. Lett.}\ }\textbf {\bibinfo
  {volume} {111}},\ \bibinfo {pages} {117202} (\bibinfo {year}
  {2013})}\BibitemShut {NoStop}%
\bibitem [{\citenamefont {Chotorlishvili}\ \emph {et~al.}(2015)\citenamefont
  {Chotorlishvili}, \citenamefont {Etesami}, \citenamefont {Berakdar},
  \citenamefont {Khomeriki},\ and\ \citenamefont {Ren}}]{PhysRevB.92.134424}%
  \BibitemOpen
  \bibfield  {author} {\bibinfo {author} {\bibfnamefont {L.}~\bibnamefont
  {Chotorlishvili}}, \bibinfo {author} {\bibfnamefont {S.~R.}\ \bibnamefont
  {Etesami}}, \bibinfo {author} {\bibfnamefont {J.}~\bibnamefont {Berakdar}},
  \bibinfo {author} {\bibfnamefont {R.}~\bibnamefont {Khomeriki}},\ and\
  \bibinfo {author} {\bibfnamefont {J.}~\bibnamefont {Ren}},\ }\bibfield
  {title} {\bibinfo {title} {Electromagnetically controlled multiferroic
  thermal diode},\ }\href {https://doi.org/10.1103/PhysRevB.92.134424}
  {\bibfield  {journal} {\bibinfo  {journal} {Phys. Rev. B}\ }\textbf {\bibinfo
  {volume} {92}},\ \bibinfo {pages} {134424} (\bibinfo {year}
  {2015})}\BibitemShut {NoStop}%
\bibitem [{\citenamefont {Lee}\ \emph {et~al.}(2010)\citenamefont {Lee},
  \citenamefont {Sai}, \citenamefont {Cai}, \citenamefont {Niu},\ and\
  \citenamefont {Demkov}}]{PhysRevB.81.144425}%
  \BibitemOpen
  \bibfield  {author} {\bibinfo {author} {\bibfnamefont {J.}~\bibnamefont
  {Lee}}, \bibinfo {author} {\bibfnamefont {N.}~\bibnamefont {Sai}}, \bibinfo
  {author} {\bibfnamefont {T.}~\bibnamefont {Cai}}, \bibinfo {author}
  {\bibfnamefont {Q.}~\bibnamefont {Niu}},\ and\ \bibinfo {author}
  {\bibfnamefont {A.~A.}\ \bibnamefont {Demkov}},\ }\bibfield  {title}
  {\bibinfo {title} {Interfacial magnetoelectric coupling in tricomponent
  superlattices},\ }\href {https://doi.org/10.1103/PhysRevB.81.144425}
  {\bibfield  {journal} {\bibinfo  {journal} {Phys. Rev. B}\ }\textbf {\bibinfo
  {volume} {81}},\ \bibinfo {pages} {144425} (\bibinfo {year}
  {2010})}\BibitemShut {NoStop}%
\bibitem [{\citenamefont {Sahoo}\ \emph {et~al.}(2007)\citenamefont {Sahoo},
  \citenamefont {Polisetty}, \citenamefont {Duan}, \citenamefont {Jaswal},
  \citenamefont {Tsymbal},\ and\ \citenamefont {Binek}}]{PhysRevB.76.092108}%
  \BibitemOpen
  \bibfield  {author} {\bibinfo {author} {\bibfnamefont {S.}~\bibnamefont
  {Sahoo}}, \bibinfo {author} {\bibfnamefont {S.}~\bibnamefont {Polisetty}},
  \bibinfo {author} {\bibfnamefont {C.-G.}\ \bibnamefont {Duan}}, \bibinfo
  {author} {\bibfnamefont {S.~S.}\ \bibnamefont {Jaswal}}, \bibinfo {author}
  {\bibfnamefont {E.~Y.}\ \bibnamefont {Tsymbal}},\ and\ \bibinfo {author}
  {\bibfnamefont {C.}~\bibnamefont {Binek}},\ }\bibfield  {title} {\bibinfo
  {title} {Ferroelectric control of magnetism in
  $\mathrm{Ba}\mathrm{Ti}\mathrm{O}_{3}\slash\mathrm{Fe}$ heterostructures via
  interface strain coupling},\ }\href
  {https://doi.org/10.1103/PhysRevB.76.092108} {\bibfield  {journal} {\bibinfo
  {journal} {Phys. Rev. B}\ }\textbf {\bibinfo {volume} {76}},\ \bibinfo
  {pages} {092108} (\bibinfo {year} {2007})}\BibitemShut {NoStop}%
\bibitem [{\citenamefont {Duan}\ \emph {et~al.}(2006)\citenamefont {Duan},
  \citenamefont {Jaswal},\ and\ \citenamefont
  {Tsymbal}}]{PhysRevLett.97.047201}%
  \BibitemOpen
  \bibfield  {author} {\bibinfo {author} {\bibfnamefont {C.-G.}\ \bibnamefont
  {Duan}}, \bibinfo {author} {\bibfnamefont {S.~S.}\ \bibnamefont {Jaswal}},\
  and\ \bibinfo {author} {\bibfnamefont {E.~Y.}\ \bibnamefont {Tsymbal}},\
  }\bibfield  {title} {\bibinfo {title} {Predicted magnetoelectric effect in
  $\mathrm{Fe}/{\mathrm{batio}}_{3}$ multilayers: Ferroelectric control of
  magnetism},\ }\href {https://doi.org/10.1103/PhysRevLett.97.047201}
  {\bibfield  {journal} {\bibinfo  {journal} {Phys. Rev. Lett.}\ }\textbf
  {\bibinfo {volume} {97}},\ \bibinfo {pages} {047201} (\bibinfo {year}
  {2006})}\BibitemShut {NoStop}%
\bibitem [{\citenamefont {Sharma}\ \emph {et~al.}(2017)\citenamefont {Sharma},
  \citenamefont {Saha}, \citenamefont {Patnaik},\ and\ \citenamefont
  {Kuanr}}]{sharma2017synthesis}%
  \BibitemOpen
  \bibfield  {author} {\bibinfo {author} {\bibfnamefont {V.}~\bibnamefont
  {Sharma}}, \bibinfo {author} {\bibfnamefont {J.}~\bibnamefont {Saha}},
  \bibinfo {author} {\bibfnamefont {S.}~\bibnamefont {Patnaik}},\ and\ \bibinfo
  {author} {\bibfnamefont {B.~K.}\ \bibnamefont {Kuanr}},\ }\bibfield  {title}
  {\bibinfo {title} {Synthesis and characterization of yttrium iron garnet
  (yig) nanoparticles-microwave material},\ }\href@noop {} {\bibfield
  {journal} {\bibinfo  {journal} {AIP Advances}\ }\textbf {\bibinfo {volume}
  {7}},\ \bibinfo {pages} {056405} (\bibinfo {year} {2017})}\BibitemShut
  {NoStop}%
\bibitem [{\citenamefont {Gibson}\ \emph {et~al.}(2020)\citenamefont {Gibson},
  \citenamefont {Bildstein}, \citenamefont {Hartman},\ and\ \citenamefont
  {Grabowski}}]{gibson2020nonlinear}%
  \BibitemOpen
  \bibfield  {author} {\bibinfo {author} {\bibfnamefont {C.}~\bibnamefont
  {Gibson}}, \bibinfo {author} {\bibfnamefont {S.}~\bibnamefont {Bildstein}},
  \bibinfo {author} {\bibfnamefont {J.~A.~L.}\ \bibnamefont {Hartman}},\ and\
  \bibinfo {author} {\bibfnamefont {M.}~\bibnamefont {Grabowski}},\ }\bibfield
  {title} {\bibinfo {title} {Nonlinear resonances and transitions to chaotic
  dynamics of a driven magnetic moment},\ }\href@noop {} {\bibfield  {journal}
  {\bibinfo  {journal} {Journal of Magnetism and Magnetic Materials}\ }\textbf
  {\bibinfo {volume} {501}},\ \bibinfo {pages} {166352} (\bibinfo {year}
  {2020})}\BibitemShut {NoStop}%
\bibitem [{\citenamefont {Gitterman}(2001)}]{gitterman2001bistable}%
  \BibitemOpen
  \bibfield  {author} {\bibinfo {author} {\bibfnamefont {M.}~\bibnamefont
  {Gitterman}},\ }\bibfield  {title} {\bibinfo {title} {Bistable oscillator
  driven by two periodic fields},\ }\href@noop {} {\bibfield  {journal}
  {\bibinfo  {journal} {Journal of Physics A: Mathematical and General}\
  }\textbf {\bibinfo {volume} {34}},\ \bibinfo {pages} {L355} (\bibinfo {year}
  {2001})}\BibitemShut {NoStop}%
\bibitem [{\citenamefont {Landa}\ and\ \citenamefont
  {McClintock}(2000)}]{landa2000vibrational}%
  \BibitemOpen
  \bibfield  {author} {\bibinfo {author} {\bibfnamefont {P.}~\bibnamefont
  {Landa}}\ and\ \bibinfo {author} {\bibfnamefont {P.~V.}\ \bibnamefont
  {McClintock}},\ }\bibfield  {title} {\bibinfo {title} {Vibrational
  resonance},\ }\href@noop {} {\bibfield  {journal} {\bibinfo  {journal}
  {Journal of Physics A: Mathematical and general}\ }\textbf {\bibinfo {volume}
  {33}},\ \bibinfo {pages} {L433} (\bibinfo {year} {2000})}\BibitemShut
  {NoStop}%
\bibitem [{\citenamefont {Zhang}\ \emph {et~al.}(2015)\citenamefont {Zhang},
  \citenamefont {Wang}, \citenamefont {Li}, \citenamefont {Luo}, \citenamefont
  {Wu}, \citenamefont {Nori},\ and\ \citenamefont {You}}]{zhang2015cavity}%
  \BibitemOpen
  \bibfield  {author} {\bibinfo {author} {\bibfnamefont {D.}~\bibnamefont
  {Zhang}}, \bibinfo {author} {\bibfnamefont {X.-M.}\ \bibnamefont {Wang}},
  \bibinfo {author} {\bibfnamefont {T.-F.}\ \bibnamefont {Li}}, \bibinfo
  {author} {\bibfnamefont {X.-Q.}\ \bibnamefont {Luo}}, \bibinfo {author}
  {\bibfnamefont {W.}~\bibnamefont {Wu}}, \bibinfo {author} {\bibfnamefont
  {F.}~\bibnamefont {Nori}},\ and\ \bibinfo {author} {\bibfnamefont
  {J.}~\bibnamefont {You}},\ }\bibfield  {title} {\bibinfo {title} {Cavity
  quantum electrodynamics with ferromagnetic magnons in a small
  yttrium-iron-garnet sphere},\ }\href@noop {} {\bibfield  {journal} {\bibinfo
  {journal} {npj Quantum Information}\ }\textbf {\bibinfo {volume} {1}},\
  \bibinfo {pages} {1} (\bibinfo {year} {2015})}\BibitemShut {NoStop}%
\bibitem [{\citenamefont {Arnol'd}(2013)}]{arnol2013mathematical}%
  \BibitemOpen
  \bibfield  {author} {\bibinfo {author} {\bibfnamefont {V.~I.}\ \bibnamefont
  {Arnol'd}},\ }\href@noop {} {\emph {\bibinfo {title} {Mathematical methods of
  classical mechanics}}},\ Vol.~\bibinfo {volume} {60}\ (\bibinfo  {publisher}
  {Springer Science \& Business Media},\ \bibinfo {year} {2013})\BibitemShut
  {NoStop}%
\bibitem [{\citenamefont {Singh}\ \emph {et~al.}(2020)\citenamefont {Singh},
  \citenamefont {Chotorlishvili}, \citenamefont {Srivastava}, \citenamefont
  {Tralle}, \citenamefont {Toklikishvili}, \citenamefont {Berakdar},\ and\
  \citenamefont {Mishra}}]{PhysRevB.101.104311}%
  \BibitemOpen
  \bibfield  {author} {\bibinfo {author} {\bibfnamefont {A.~K.}\ \bibnamefont
  {Singh}}, \bibinfo {author} {\bibfnamefont {L.}~\bibnamefont
  {Chotorlishvili}}, \bibinfo {author} {\bibfnamefont {S.}~\bibnamefont
  {Srivastava}}, \bibinfo {author} {\bibfnamefont {I.}~\bibnamefont {Tralle}},
  \bibinfo {author} {\bibfnamefont {Z.}~\bibnamefont {Toklikishvili}}, \bibinfo
  {author} {\bibfnamefont {J.}~\bibnamefont {Berakdar}},\ and\ \bibinfo
  {author} {\bibfnamefont {S.~K.}\ \bibnamefont {Mishra}},\ }\bibfield  {title}
  {\bibinfo {title} {Generation of coherence in an exactly solvable nonlinear
  nanomechanical system},\ }\href {https://doi.org/10.1103/PhysRevB.101.104311}
  {\bibfield  {journal} {\bibinfo  {journal} {Phys. Rev. B}\ }\textbf {\bibinfo
  {volume} {101}},\ \bibinfo {pages} {104311} (\bibinfo {year}
  {2020})}\BibitemShut {NoStop}%
\bibitem [{\citenamefont {Singh}\ \emph {et~al.}(2022)\citenamefont {Singh},
  \citenamefont {Chotorlishvili}, \citenamefont {Toklikishvili}, \citenamefont
  {Tralle},\ and\ \citenamefont {Mishra}}]{singh2022hybrid}%
  \BibitemOpen
  \bibfield  {author} {\bibinfo {author} {\bibfnamefont {A.}~\bibnamefont
  {Singh}}, \bibinfo {author} {\bibfnamefont {L.}~\bibnamefont
  {Chotorlishvili}}, \bibinfo {author} {\bibfnamefont {Z.}~\bibnamefont
  {Toklikishvili}}, \bibinfo {author} {\bibfnamefont {I.}~\bibnamefont
  {Tralle}},\ and\ \bibinfo {author} {\bibfnamefont {S.}~\bibnamefont
  {Mishra}},\ }\bibfield  {title} {\bibinfo {title} {Hybrid quantum--classical
  chaotic nems},\ }\href@noop {} {\bibfield  {journal} {\bibinfo  {journal}
  {Physica D: Nonlinear Phenomena}\ }\textbf {\bibinfo {volume} {439}},\
  \bibinfo {pages} {133418} (\bibinfo {year} {2022})}\BibitemShut {NoStop}%
\bibitem [{\citenamefont {Zaslavsky}(2007)}]{zaslavsky2007physics}%
  \BibitemOpen
  \bibfield  {author} {\bibinfo {author} {\bibfnamefont {G.~M.}\ \bibnamefont
  {Zaslavsky}},\ }\href@noop {} {\emph {\bibinfo {title} {The physics of chaos
  in Hamiltonian systems}}}\ (\bibinfo  {publisher} {world scientific},\
  \bibinfo {year} {2007})\BibitemShut {NoStop}%
\bibitem [{\citenamefont {Greenspan}\ and\ \citenamefont
  {Holmes}(1984)}]{greenspan1984repeated}%
  \BibitemOpen
  \bibfield  {author} {\bibinfo {author} {\bibfnamefont {B.}~\bibnamefont
  {Greenspan}}\ and\ \bibinfo {author} {\bibfnamefont {P.}~\bibnamefont
  {Holmes}},\ }\bibfield  {title} {\bibinfo {title} {Repeated resonance and
  homoclinic bifurcation in a periodically forced family of oscillators},\
  }\href@noop {} {\bibfield  {journal} {\bibinfo  {journal} {SIAM journal on
  mathematical analysis}\ }\textbf {\bibinfo {volume} {15}},\ \bibinfo {pages}
  {69} (\bibinfo {year} {1984})}\BibitemShut {NoStop}%
\bibitem [{\citenamefont {Yamada}\ \emph {et~al.}(2020)\citenamefont {Yamada},
  \citenamefont {Kogiso}, \citenamefont {Shiota}, \citenamefont {Yamamoto},
  \citenamefont {Yamaguchi}, \citenamefont {Moriyama}, \citenamefont {Ono},\
  and\ \citenamefont {Shima}}]{yamada2020dependence}%
  \BibitemOpen
  \bibfield  {author} {\bibinfo {author} {\bibfnamefont {K.}~\bibnamefont
  {Yamada}}, \bibinfo {author} {\bibfnamefont {K.}~\bibnamefont {Kogiso}},
  \bibinfo {author} {\bibfnamefont {Y.}~\bibnamefont {Shiota}}, \bibinfo
  {author} {\bibfnamefont {M.}~\bibnamefont {Yamamoto}}, \bibinfo {author}
  {\bibfnamefont {A.}~\bibnamefont {Yamaguchi}}, \bibinfo {author}
  {\bibfnamefont {T.}~\bibnamefont {Moriyama}}, \bibinfo {author}
  {\bibfnamefont {T.}~\bibnamefont {Ono}},\ and\ \bibinfo {author}
  {\bibfnamefont {M.}~\bibnamefont {Shima}},\ }\bibfield  {title} {\bibinfo
  {title} {Dependence of gilbert damping constant on microstructure in
  nanocrystalline yig coatings prepared by co-precipitation and spin-coating on
  a si substrate},\ }\href@noop {} {\bibfield  {journal} {\bibinfo  {journal}
  {Journal of Magnetism and Magnetic Materials}\ }\textbf {\bibinfo {volume}
  {513}},\ \bibinfo {pages} {167253} (\bibinfo {year} {2020})}\BibitemShut
  {NoStop}%
\bibitem [{\citenamefont {Chotorlishvili}\ \emph {et~al.}(2011)\citenamefont
  {Chotorlishvili}, \citenamefont {Ugulava}, \citenamefont {Mchedlishvili},
  \citenamefont {Komnik}, \citenamefont {Wimberger},\ and\ \citenamefont
  {Berakdar}}]{chotorlishvili2011nonlinear}%
  \BibitemOpen
  \bibfield  {author} {\bibinfo {author} {\bibfnamefont {L.}~\bibnamefont
  {Chotorlishvili}}, \bibinfo {author} {\bibfnamefont {A.}~\bibnamefont
  {Ugulava}}, \bibinfo {author} {\bibfnamefont {G.}~\bibnamefont
  {Mchedlishvili}}, \bibinfo {author} {\bibfnamefont {A.}~\bibnamefont
  {Komnik}}, \bibinfo {author} {\bibfnamefont {S.}~\bibnamefont {Wimberger}},\
  and\ \bibinfo {author} {\bibfnamefont {J.}~\bibnamefont {Berakdar}},\
  }\bibfield  {title} {\bibinfo {title} {Nonlinear dynamics of two coupled
  nano-electromechanical resonators},\ }\href@noop {} {\bibfield  {journal}
  {\bibinfo  {journal} {Journal of Physics B: Atomic, Molecular and Optical
  Physics}\ }\textbf {\bibinfo {volume} {44}},\ \bibinfo {pages} {215402}
  (\bibinfo {year} {2011})}\BibitemShut {NoStop}%
\bibitem [{\citenamefont {Bogoliubov}\ and\ \citenamefont
  {Mitropolski}(1961)}]{bogoliubov1961asymptotic}%
  \BibitemOpen
  \bibfield  {author} {\bibinfo {author} {\bibfnamefont {N.~N.}\ \bibnamefont
  {Bogoliubov}}\ and\ \bibinfo {author} {\bibfnamefont {Y.~A.}\ \bibnamefont
  {Mitropolski}},\ }\bibfield  {title} {\bibinfo {title} {Asymptotic methods in
  the theory of non-linear oscillations},\ }\href@noop {} {\bibfield  {journal}
  {\bibinfo  {journal} {Asymptotic Methods in the Theory of Non-Linear
  Oscillations}\ } (\bibinfo {year} {1961})}\BibitemShut {NoStop}%
\bibitem [{\citenamefont {Ugulava}\ \emph {et~al.}(2005)\citenamefont
  {Ugulava}, \citenamefont {Chotorlishvili},\ and\ \citenamefont
  {Nickoladze}}]{PhysRevE.71.056211}%
  \BibitemOpen
  \bibfield  {author} {\bibinfo {author} {\bibfnamefont {A.}~\bibnamefont
  {Ugulava}}, \bibinfo {author} {\bibfnamefont {L.}~\bibnamefont
  {Chotorlishvili}},\ and\ \bibinfo {author} {\bibfnamefont {K.}~\bibnamefont
  {Nickoladze}},\ }\bibfield  {title} {\bibinfo {title} {Irreversible evolution
  of quantum chaos},\ }\href {https://doi.org/10.1103/PhysRevE.71.056211}
  {\bibfield  {journal} {\bibinfo  {journal} {Phys. Rev. E}\ }\textbf {\bibinfo
  {volume} {71}},\ \bibinfo {pages} {056211} (\bibinfo {year}
  {2005})}\BibitemShut {NoStop}%
\bibitem [{\citenamefont {Grassberger}\ and\ \citenamefont
  {Procaccia}(1983)}]{PhysRevLett.50.346}%
  \BibitemOpen
  \bibfield  {author} {\bibinfo {author} {\bibfnamefont {P.}~\bibnamefont
  {Grassberger}}\ and\ \bibinfo {author} {\bibfnamefont {I.}~\bibnamefont
  {Procaccia}},\ }\bibfield  {title} {\bibinfo {title} {Characterization of
  strange attractors},\ }\href {https://doi.org/10.1103/PhysRevLett.50.346}
  {\bibfield  {journal} {\bibinfo  {journal} {Phys. Rev. Lett.}\ }\textbf
  {\bibinfo {volume} {50}},\ \bibinfo {pages} {346} (\bibinfo {year}
  {1983})}\BibitemShut {NoStop}%
\end{thebibliography}%

\end{document}